\documentclass[preprint,9pt]{elsarticle}
\usepackage{amssymb, stmaryrd}
\usepackage{graphicx}
\usepackage{slashbox}
\usepackage{subfigure}
\usepackage{hyperref}

\journal{Journal of Computer and System Sciences}

\def\>{\ensuremath{\rangle}}
\def\<{\ensuremath{\langle}}

\def\-{\ensuremath{\textrm{-}}}

\def\fdmu{\Delta}

\def\fpi{\widehat{\pi}}
\def\h{\ensuremath{\mathcal{H}}}

\def\g{\ensuremath{\mathcal{G}}}
\def\lh{\ensuremath{\mathcal{L(H)}}}
\def\dh{\ensuremath{\mathcal{D(H})}}
\def\q{\bold Q}
\def\Q{\ensuremath{\mathbb Q}}
\def\P{\ensuremath{\mathbb P}}
\def\SO{\ensuremath{\mathcal{S}}}
\def\HP{\ensuremath{\mathcal{HP}}}
\def\hpe{\ensuremath{\mathcal{\e}}}

\def\r{\ensuremath{\mathcal{R}}}

\def\m{\ensuremath{\mathcal{M}}}

\def\s{\ensuremath{\mathcal{S}}}
\def\t{\ensuremath{\mathcal{T}}}

\def\U{\ensuremath{\mathfrak{U}}}

\def\x{\ensuremath{\mathcal{X}}}
\def\y{\ensuremath{\mathcal{Y}}}
\def\z{\ensuremath{\mathcal{Z}}}

\def\ra{\ensuremath{\rightarrow}}
\def\a{\ensuremath{\mathcal{E}}}
\def\b{\ensuremath{\mathcal{F}}}
\def\c{\ensuremath{\mathcal{G}}}

\def\e{\ensuremath{\mathcal{E}}}
\def\f{\ensuremath{\mathcal{F}}}

\def\X{\mbox{\bf{X}}}
\def\U{\mbox{\bf{U}}}
\def\N{\mathbb{N}}
\def\sreal{\mathbb{R}}

\newcommand {\while} {\mbox{\bf{while}}}
\newcommand {\ddo} {\mbox{\bf{do}}}

\def\dh{\ensuremath{\mathcal{D(H)}}}
\def\lh{\ensuremath{\mathcal{L(H)}}}
\def\le{\ensuremath{\sqsubseteq}}
\def\ge{\ensuremath{\sqsupseteq}}

\def\leqI{\ensuremath{\mathcal{S}^1(\h)}}

\newcommand {\iif} {\mbox{\bf{if}}}
\newcommand {\then} {\mbox{\bf{then}}}

\newcommand {\true} {\mbox{\texttt{tt}}}

\newcommand{\tr}{{\rm tr}}

\newcommand{\id}{\mathcal{I}}

\newtheorem{theorem}{Theorem}[section]
\newtheorem{example}[theorem]{Example}
\newtheorem{lemma}[theorem]{Lemma}
\newtheorem{definition}[theorem]{Definition}

\newtheorem{corollary}[theorem]{Corollary}
\begin{document}

\begin{frontmatter}

\title{Model checking quantum Markov chains}


%
%
\author{Yuan Feng, Nengkun Yu, and Mingsheng Ying}
%

\address{University of Technology, Sydney, Australia, \\
Tsinghua University, China
}
%
%

\begin{abstract} Although security of quantum cryptography is provable based on principles of quantum mechanics, it can be compromised by flaws in the design of quantum protocols.
So, it is indispensable to develop techniques for verifying and debugging quantum cryptographic systems. Model-checking has proved to be effective in the verification of classical cryptographic protocols, but an essential difficulty arises when it is applied to quantum systems: the state space of a quantum system is always a continuum even when its dimension is finite. To overcome this difficulty, we introduce a novel notion of \emph{quantum Markov chain}, especially suited for modelling quantum cryptographic protocols, in which quantum effects are encoded as super-operators labelling transitions, leaving the location information (nodes) being classical. Then we define a quantum extension of probabilistic computation tree logic (PCTL) and develop a model-checking algorithm for quantum Markov chains.       
\end{abstract}

\end{frontmatter}

\section{Introduction}

Quantum cryptography,  which uses quantum mechanical effects to accomplish cryptographic tasks, has been developed so rapidly that quantum cryptographic systems
are already commercially available by a number of companies such as Id Quantique, MagiQ Technologies, QuintessenceLabs, and NEC~\cite{companies}. One of the greatest advantages of quantum cryptography over its classical counterpart 
is that the security and ability to detect the presence of eavesdropping are provable based on principles of
quantum mechanics. In practice, however, errors which comprise this $absolute$ security may still creep in the protocol design level: As quantum mechanics is counter-intuitive, quantum cryptographic protocol designers
will inevitably make more mistakes than classical protocol designers, especially when more and more complicated quantum
protocols can be implemented by future physical technology.
Therefore, it is indispensable to develop methodologies and techniques for the verification of quantum cryptographic systems. 

Over the last four decades, model-checking~\cite{CGP99,BK08} has become one of the dominant techniques for verification
of classical hardware as well as software systems, and has proved mature as witnessed by a large number of successful industrial applications. Model-checking techniques have also been widely used in the verification of security protocols~\cite{Lo96}.
One of the advantages of model checking is that it is usually automatic and provides counter-examples, which are indispensable in debugging, in case 
the property is violated. 

Given its advantage stated above, people started to explore the possibility of applying model-checking in the verification of quantum cryptographic protocols. The main obstacle for model checking quantum systems is 
that the set of all quantum states, traditionally regarded as the underlying state space of the models to be checked, is a continuum. Hence the techniques of classical model checking, which normally work only for a finite state space, cannot be applied directly. Gay et al.~\cite{GNP06,Pa08} provided a clever solution for this problem by restricting the state space to a set of finitely describable states called stabiliser states, and restricting the quantum operations applied on them to the class of Clifford group. By doing this, they were able to obtain an efficient model checker for quantum protocols, employing purely classical algorithms. 
They even developed an automatic tool QMC (Quantum Model-Checker) for model-checking quantum communication protocols~\cite{GNP08}. 
However, the limitation of their approach is also obvious: it can only check the (partial) behaviours of a protocol on stabiliser states, and
does not work for more general protocols. A similar idea was independently introduced by Hung et al.~\cite{HSY+04,WS06} to synthesise quantum circuits. By formulating quantum logic synthesis problem via symbolic reachability analysis, they were able to reduce the original problem to multiple-valued logic synthesis, thus simplifying the search space and algorithm complexity.

This paper presents another solution to the problem, which applies to general quantum protocols. 
We propose a novel notion of quantum Markov chain where quantum effects are entirely encoded into super-operators labelling transitions, and the nodes of its transition graph carry only classical information and thus they are discrete. In this way, the state spaces of quantum Markov chains become countable, and often finite. 
However, the following new difficulty has to be overcome, namely,   

\begin{center}
\begin{minipage}[h]{0.9\linewidth}
A prerequisite for defining probabilistic temporal logic is a suitable probability measure on the set of infinite paths of a Markov chain. Vardi~\cite{Var85} introduced such a measure by letting the $\sigma$-algebra be generated by cylinder extensions of finite paths and proved that the events of infinite paths specified by various temporal logical formulas are measurable. The probabilities of these cylinder sets are given in a natural way. Then the probability measure on the cylinder sets can be extended to Vardi's $\sigma$-algebra by the Carath\'eodory-Hahn extension theorem. For a quantum Markov chain, however, a super-operator valued measure instead of a numerical measure must be introduced because its transitions are labelled by super-operators instead of numerical probabilities. How can we apply Vardi's procedure to this new kind of measures?    
\end{minipage}
\end{center}

This paper solves the above problem by employing Kluvanek's generalisation of the Carath\'eodory-Hahn extension theorem from vector measure theory~\cite{DU77}. Furthermore, we define a quantum extension of PCTL and develop an algorithm for model-checking quantum Markov chains. In particular, a large part of classical techniques are adapted to verify properties of quantum systems expressed in this logic.

We assume the readers are familiar with the basic notions of 
linear algebra and quantum theory. We put a brief introduction into the appendix for the convenience of the readers.
For more details, we refer to \cite{NC00}.

\subsection{Related work}

The mathematical structure employed in this paper to model quantum systems is a super-operator weighted quantum Markov chain. The idea of defining the denotational semantics of a quantum program as a super-operator was first proposed by Selinger~\cite{Se04}. Prior to our work, there were quite a few different notions of quantum Markov chains~\cite{Ac76,BP02,FS10,YYFD12}, introduced by authors from different research communities. The major difference is that in their models transitions are considered between quantum states which always form a continuum, whereas  in our model transitions are considered between different points in an execution path, and quantum operations are treated as labels of the transitions. Consequently, the state spaces of our quantum Markov chains are typically finite, and classical model checking techniques can be easily adapted to verify quantum systems. 

A quantum Markov model similar to that used in this paper was introduced by Gudder~\cite{Gu08} where the transition between different vertices is characterised by an operation matrix whose entries are super-operators and the sum of each column is trace-preserving. However, the motivations are very different: Gudder~\cite{Gu08} aimed at defining a pure mathematical generalisation of quantum walks, whereas our model is extracted from semantics of quantum protocols and quantum programs.

An exogenous quantum computation tree logic has already been proposed in~\cite{BCM08}, which is very powerful and can express quantum states in a Hilbert space as well as quantum operations performed on them.
As a result, it can be used for reasoning about evolution of quantum (software as well as hardware) systems.
The QCTL presented in this paper, however, only consider \emph{classical} properties as its atomic propositions. This is in accordance with our notion of quantum Markov chains where the state space is classical. Thus our approach is suitable for model checking the classical aspect of quantum \emph{software} systems. Note that a large part of quantum communication protocols, such as superdense coding \cite{BW92}, BB84 quantum key distribution \cite{BB84}, quantum leader election \cite{Tani:2012es} et al, all aim at achieving some classical tasks. This is not a very serious limitation.  

\section{Super-operators and super-operator valued measures} \label{sec:so}

For the sake of simplicity, in the following we use the term super-operator to denote a completely positive super-operator.
Let $\SO(\h)$ be the set of super-operators on $\h$, ranged over by $\a, \b, \cdots$.
Obviously, both $(\SO(\h), 0_\h, +)$ and $(\SO(\h), \id_\h, \circ)$ are monoids, where $\id_\h$ and $0_\h$ are the identity and null super-operators on $\h$, respectively, and $\circ$ is the composition of super-operators defined by $(\a\circ\b)(\rho) = \a(\b(\rho))$ for any $\rho\in \dh$ where $\dh$ is the set of density operators on $\h$. We always omit the symbol $\circ$ and write $\a\b$ directly
for $\a\circ \b$. Furthermore, the operation $\circ$ is (both left and right) distributive with respect to $+$: $$\a(\b_1+\b_2)=\a\b_1 + \a\b_2, \ \ (\b_1+\b_2)\a=\b_1\a + \b_2\a.$$ Thus $(\SO(\h), +, \circ)$ forms a semiring.
We will use two different orders:
\begin{definition}\label{def:orders} Let $\a, \b\in \SO(\h)$. 
\begin{enumerate}
\item $\a\le \b$ if for any $\rho\in \dh$, $\b(\rho) - \a(\rho)$ is positive semi-definite;
\item $\a\lesssim \b$ if for any $\rho\in \dh$, $\tr(\a(\rho))\leq \tr(\b(\rho))$.
\end{enumerate}
\end{definition}
The first order is lifted from L\"{o}wner partial order on density operators, whereas the second one is used to compare the ability of `trace preservation'. 
Note that the trace of a (unnormalised) quantum state is exactly the probability that the (normalised) state is reached~\cite{Se04}.
Intuitively, $\a\lesssim \b$ if and only if the success probability of performing $\e$ is always not greater than that of performing $\f$, whatever the initial state is.

Note that $\le$ is a partial order while $\lesssim$ is a pre-order. 
Let $\eqsim$ be $\lesssim \cap \gtrsim$; it is obviously an equivalence relation.
\begin{lemma}\label{lem:ordersrel} Let $\a, \b\in \SO(\h).$ Then
\begin{enumerate}
\item $\a\le\b$ implies $\a\lesssim\b$, while the reverse is not true in general. 
\item $\a\lesssim\b$ implies $\a\le \b'$ for some $\b'\eqsim\b$.
\end{enumerate}
\end{lemma}
{\it Proof.}
(1) is obvious. To prove (2), let $\a=\{ E_i : i\in I\}$ and $\b=\{ F_j : j\in J\}$ be the Kraus representation of $\a$ and $\b$, respectively. Then $\a\lesssim \b$ if and only if $G=\sum_{j\in J} F_j^\dag F_j-\sum_{i\in I}  E_i^\dag E_i$ is positive semi-definite. Let $G=E^\dag E$ and define the super-operator $\g = \{E\}$. Let $\b'=\a+\g$. Then it is easy to check $\b'\eqsim\b$ and $\a\le \b'$. 
\qed

The next lemma shows that the two orders $\lesssim$ and $\le$ are preserved by the right application of composition. 
\begin{lemma}\label{lem:rightapp} Let $\a,\b,\c\in \SO(\h)$. If $\a\lesssim \b$, then $\a\c\lesssim\b\c$, and
if $\a\le \b$, then $\a\c\le\b\c$.
\end{lemma}
Let $$\leqI = \{\a\in \SO(\h) : \a\lesssim\id_\h\}.$$
Let
$\leqI^n$ 
be the set of $n$-size row vectors over $\leqI$, and extend the partial order $\le$ componentwise to it. Then we have
\begin{lemma}\label{lem:cpov}
The set $(\leqI^n, \le)$ is a complete partial order set with the least element $(0_\h, \dots, 0_\h)$.
\end{lemma} 
{\it Proof.}
The case when $n=1$ is from~\cite{Se04}, while the extension to $n>1$ is obvious.
\qed

With the notations and properties presented above, we can prove the main result of this section, which is the key to verifying the long-run properties of quantum Markov chains. To make the paper more readable, we present the proof in  Appendix~\ref{app:2.5}.
\begin{theorem}\label{thm:fixpoint} 
Let $\t$ and $\widetilde{\g}$ be two matrices of  super-operators with sizes $n\times n$ and $1\times n$, respectively, and for each $j$, $\sum_i \t_{i,j} + \widetilde{\g}_j \lesssim \id_\h$. Let
\begin{equation}\label{eq:les}
f(X) = X \t + \widetilde{\g}
\end{equation} be a function from $\SO(\h)^n$ to $\SO(\h)^n$. Then
\begin{enumerate}
\item $f(X)$ has the least fixed point, denoted by $\widetilde{\e}$, in $\leqI^n$ with respect to the order $\le$; 
\item Given any $\e\in \leqI$ and $1\leq i\leq n$, it can be decided whether $\e\sim \widetilde{\e}_i$, $\sim{\in}\{\lesssim, \gtrsim\}$, in time $O(n^2d^4)$ where $d=dim(\h)$ is the dimension of $\h$.
\end{enumerate}
\end{theorem}

To conclude this section, we introduce the notion of super-operator valued measures, which will play a role similar to probability measures for probabilistic model checking.

\begin{definition}
Let $(\Omega, \Sigma)$ be a measurable space; that is, $\Omega$ is a non-empty set and $\Sigma$ a $\sigma$-algebra over $\Omega$. A function
$\fdmu : \Sigma \ra \leqI$ is said to be a super-operator valued measure (SVM for short) if $\fdmu$ satisfies the following properties:
\begin{enumerate}
\item $\fdmu(\Omega) \eqsim \id_\h$;
\item $\fdmu(\biguplus_i A_i) \eqsim  \sum_i\fdmu(A_i)$ for any pairwise disjoint and countable sequence $A_1$, $A_2$, $\dots$ in $\Omega$. 
\end{enumerate}
We call the triple $(\Omega, \Sigma, \fdmu)$ a (super-operator valued) measure space.
\end{definition}

We write $A=\biguplus_{i} A_i$ if $A=\bigcup_{i} A_i$ and $A_i$s are pairwise disjoint; that is, for any $i\neq j$, $A_i\cap A_j=\emptyset$.
SVMs enjoy some similar properties satisfied by probabilistic measures, which are collected as follows.
\begin{lemma}\label{lem:distprop}
 Let $(\Omega, \Sigma, \fdmu)$ be a measure space. Then
\begin{enumerate}
\item $\fdmu(\emptyset) = 0_\h$;
\item $\fdmu(A^c)  + \fdmu(A)\eqsim \id_\h$ where $A^c$ is the complement set of $A$ in $\Omega$;
\item (monotonicity) for any $A, A'\in \Sigma$, if $A\subseteq A'$ then $\fdmu(A)\lesssim \fdmu(A')$;
\item (continuity) for any sequence $A_1, A_2, \dots$ in $\Sigma$, 
\begin{itemize}
\item if $A_1\subseteq A_2\subseteq \dots$, then there exists a sequence $\e_1\le \e_2\le \dots$ in $\leqI$ such that for any $i$, $\fdmu(A_i)\eqsim \e_i$, and
$\fdmu(\bigcup_{i\geq 1} A_i)\eqsim \lim_{i\ra \infty} \e_i.$
\item if $A_1\supseteq A_2\supseteq \dots$, then there exists a sequence $\e_1\ge \e_2\ge \dots$ in $\leqI$ such that for any $i$, $\fdmu(A_i)\eqsim \e_i$, and
$\fdmu(\bigcap_{i\geq 1} A_i)\eqsim \lim_{i\ra \infty} \e_i.$
\end{itemize}
\end{enumerate}
\end{lemma}
{\it Proof.} We only prove the first item of (4). Suppose $A_1\subseteq A_2\subseteq \dots$. Let $B_n = A_n\backslash \bigcup_{i<n} A_i$, $n=1,2,\dots$.
Then each pair $B_i$ and $B_j$ are disjoint provided that $i\neq j$, and for each $n$, $A_n=\biguplus_{i\leq n} B_i$. Let $\e_n = \sum_{i\leq n} \fdmu(B_i)$. Then $\e_1\le \e_2\le \dots$, and by the additivity of $\fdmu$ we have
$\fdmu(A_n)\eqsim \e_n$. Finally,
$$\fdmu(\bigcup_{i\geq 1} A_i) = \fdmu(\biguplus_{i\geq 1} B_i)\eqsim \sum_{i\geq 1} \fdmu(B_i) = \lim_{n\ra\infty} \e_n.$$
Here the existence of the limit is guaranteed by Lemma~\ref{lem:cpov}.
\qed

\section{Super-operator weighted Markov chains}
We now extend classical Markov chains to super-operator weighted
ones. 

\begin{definition}
A super-operator weighted Markov chain, or quantum Markov chain (qMC), is a tuple $(S, \q,  AP, L)$, where
\begin{enumerate}
\item $S$ is a countable (typically finite) set of states;
\item $\q : S\times S\ra \leqI$ is called the transition matrix where for each $s\in S$, $\sum_{t\in S}\q(s, t)\eqsim \id_\h$;
\item $AP$ is a finite set of atomic propositions;
\item $L$ is a mapping from $S$ to $2^{AP}$.
\end{enumerate}
\end{definition}
A classical Markov chain may be viewed as a degenerate quantum Markov chain in which
all super-operators in the transition matrix have the form $p\id_\h$ for some $0\leq p \leq 1$. 
Let $\m=(S, \q, AP, L)$ be a quantum Markov chain. A path $\pi$ of $\m$ is an infinite sequence of states
$s_0s_1\dots$ where for all $i\geq 0$, $s_i\in S$ and $\q(s_i, s_{i+1})\neq 0_\h$. A finite path $\widehat{\pi}$ is a finite-length prefix 
of a path, and its length, denoted $|\widehat{\pi}|$, is defined to be the number of states in it. We denote by $\pi(i)$ the $i$th state of a path $\pi$,
and $\widehat{\pi}(i)$ the $i$th state of a finite path $\widehat{\pi}$ if $i<|\widehat{\pi}|$. Note that we index the states in a path or finite path from 0.
The sets of all infinite and finite paths of $\m$ starting in state $s$ are denoted $\mathit{Path^{\m}(s)}$ and 
$\mathit{Path_{fin}^{\m}(s)}$, respectively.

In order to reason about the behaviour of a qMC, we need to determine the accumulated super-operator along certain paths. 
To this end, we construct a SVM $Q_s$ for each $s\in S$ as follows. For any finite path
$\widehat{\pi}=s_0\dots s_n\in \mathit{Path_{fin}^{\m}(s)}$, we define the super-operator
$$\q(\widehat{\pi}) = 
\left\{
\begin{array}{ll}
\id_\h,  & \mbox{if } n=0;  \\
\q(s_{n-1}, s_{n})\cdots \q(s_0, s_1), & \mbox{otherwise}.
\end{array}
\right.
$$
Next we define the cylinder set $Cyl(\widehat{\pi})\subseteq \mathit{Path^{\m}(s)}$ as
$$Cyl(\widehat{\pi})  = \{\pi\in \mathit{Path^{\m}(s)} : \widehat{\pi} \mbox{ is a prefix of }\pi\};$$
that is, the set of all infinite paths with prefix $\widehat{\pi}$.
It is easy to check that the set
$$\s^\m(s) = \{Cyl(\widehat{\pi}) : \widehat{\pi}\in \mathit{Path_{fin}^{\m}(s)}\}\cup \{\emptyset\}$$
is a semi-algebra on $\mathit{Path^{\m}(s)}$. Let a mapping $Q_s$ from $\s^\m(s)$ to $\leqI$ be defined by letting $Q_s(\emptyset)=0_\h$ and 
\begin{equation}\label{eq:qs}
Q_s(Cyl(\widehat{\pi}))=\q(\widehat{\pi}).
\end{equation} 

The following theorem states that we can extend the mapping $Q_s$ defined above to a super-operator valued measure on
the $\sigma$-algebra of infinite paths generated by cylinder extensions of finite paths, similar to Vardi's
work for classical Markov chains~\cite{Var85}. The main tool we use here is Kluvanek's generalisation of the Carath\'eodory-Hahn extension theorem from vector measure theory~\cite{DU77}. We defer the detailed proof to Appendix~\ref{sec:pmain} for the sake of readability.

\begin{theorem}\label{thm:mainthm} The mapping $Q_s$ defined above can be extended to a SVM, denoted by $Q_s$ again, on the $\sigma$-algebra generated by 
$\s^\m(s)$. Furthermore, this extension is unique up to the equivalence relation $\eqsim$.
\end{theorem}

\begin{figure}[t]
\[\begin{array}{ll}
\includegraphics[width=0.3\textwidth]{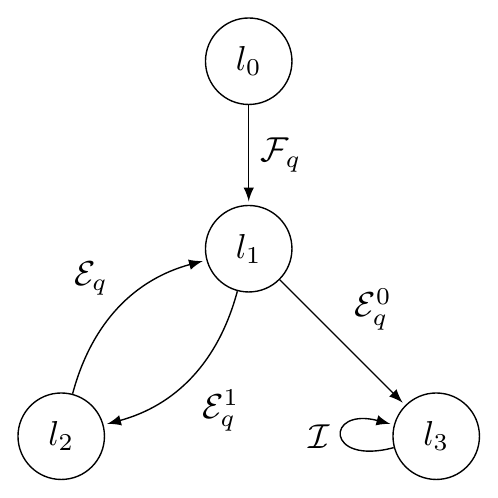}\ \ 
&\ \ \includegraphics[height=0.38\textheight]{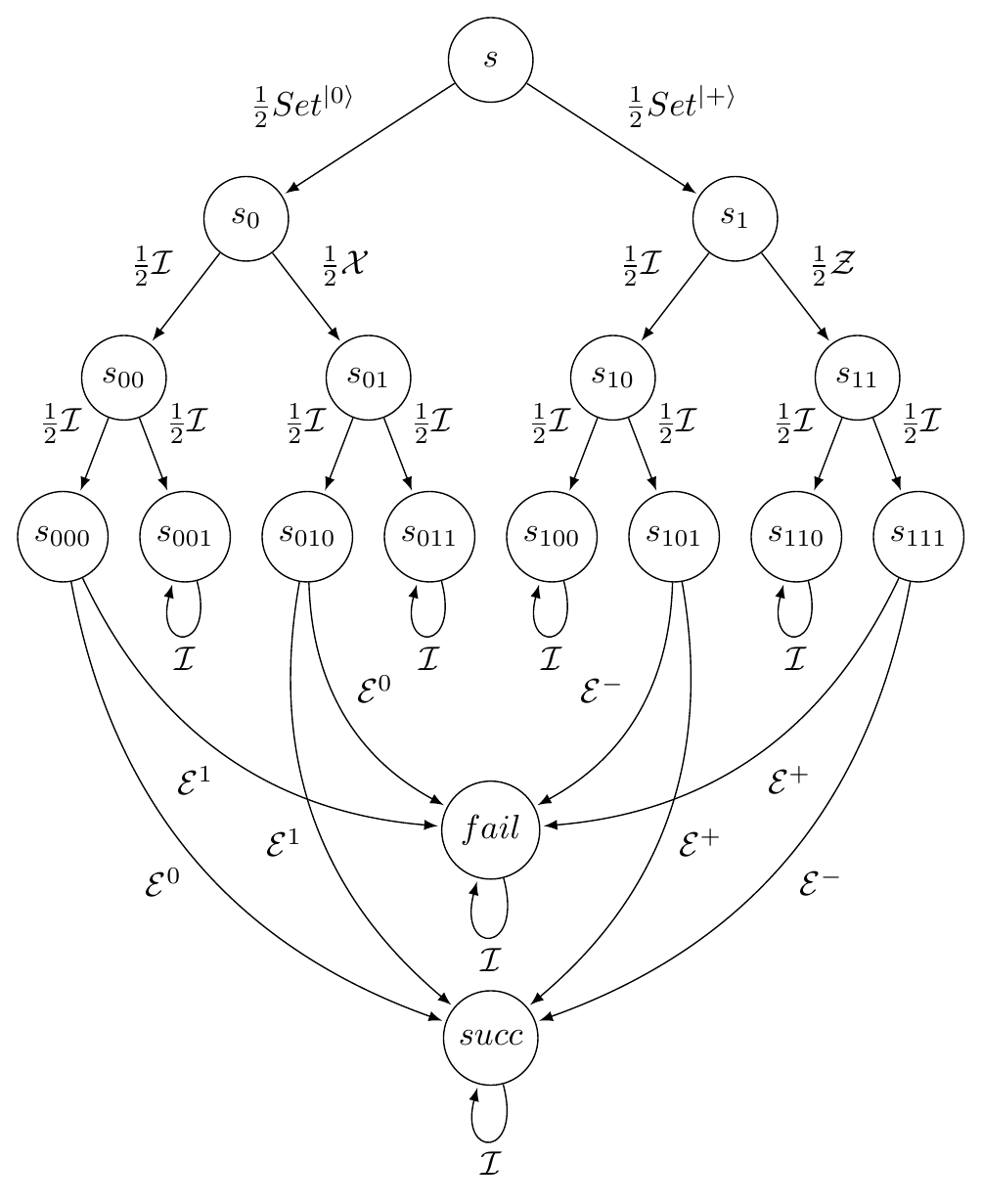}
\end{array}
 \]\caption{qMCs for a quantum loop grogram (left) and BB84 protocol when $n=1$ (right).\label{fig:bb84}}
\end{figure}

To show the expressiveness of quantum Markov chains, we present some examples.
\begin{example}\label{exm:qprog} \rm (quantum loop programs)
A simple quantum loop program goes as follows:
\begin{eqnarray*}
l_0& : & q := \f(q)\\
l_1& : & \while\ M[q]\ \ddo\\
l_2& : & \hspace{2em} q := \e(q) \\
l_3& : & \bold{od}
\end{eqnarray*} where $M=\lambda_0 |0\>\<0| + \lambda_1 |1\>\<1|$.
The intuitive meaning of this program is as follows. We first initialise the state of the quantum system $q$
at line $l_0$ by a trace-preserving super-operator $\f$. At line $l_1$, the two-outcome projective measurement $M$ is applied to
$q$. If the outcome $\lambda_0$ is observed, then the program terminates at line $l_3$; otherwise it proceeds
to $l_2$ where a trace-preserving super-operator $\e$ is performed at $q$, and then the program returns to line $l_1$ and another iteration continues.

We now construct a qMC to describe the program. Let $S=\{l_i : 0\leq i\leq 3\}$,
$AP=S$, $L(l_i)=\{l_i\}$ for each $i$, and $\q$ be defined as 
$\q(l_0, l_1) = \f_q$,  $\q(l_1, l_3) = \e^0_q=\{|0\>_q\<0|\}$, 
$\q(l_1, l_2) =  \e^1_q=\{|1\>_q\<1|\}$, 
$\q(l_2, l_1) = \e_q$,  and $\q(l_3, l_3) = \id_\h$. The qMC is depicted in Figure~\ref{fig:bb84} (left).
Let $\widehat{\pi}=l_0l_1l_2l_1l_2l_1l_3$ be a finite path from $l_0$. Then
$$\q(\widehat{\pi}) =\q(l_1, l_3)\q(l_2, l_1)\q(l_1, l_2)\q(l_2, l_1)\q(l_1, l_2)\q(l_0, l_1)=\e^0_q\e_q\e^1_q\e_q\e^1_q\f_q.$$
\end{example}

\begin{example}\label{exm:bb84}\rm (Quantum key-distribution)
BB84, the first quantum key distribution protocol developed by Bennett and Brassard in 1984 \cite{BB84}, provides a provably secure way to
create a private key between two parties, say, Alice and Bob.
The basic BB84 protocol goes as follows:
\begin{enumerate}
\item Alice randomly creates two strings of bits $\tilde{B}_a$ and  $\tilde{K}_a$, each with size $n$.
\item Alice prepares a string of qubits $\tilde{q}$, with size $n$, such that 
the $i$th qubit of $\tilde{q}$ is $|x_y\>$ where $x$ and $y$ are the $i$th bits of $\tilde{B}_a$ and  $\tilde{K}_a$, respectively,
and $|0_0\> = |0\>$, $|0_1\> = |1\>$,
$|1_0\> = |+\>= (|0\> + |1\>)/\sqrt{2}$, and $|1_1\> = |-\>= (|0\> - |1\>)/\sqrt{2}$. 
\item Alice sends the qubit string $\tilde{q}$ to Bob.
\item Bob randomly generates a string of bits $\tilde{B}_b$ with size $n$.
\item Bob measures each qubit received from Alice according to the basis determined by the bits he generated: if the $i$th bit of $\tilde{B}_b$ is $k$ then
he measures the $i$th qubit of $\tilde{q}$ with $\{|k_0\>, |k_1\>\}$, $k=0,1$. Let the measurement results be $\tilde{K}_b$, which is also a string of bits with size $n$.
\item Bob sends his choice of measurement bases $\tilde{B}_b$ back to Alice, and upon receiving the information, Alice sends her bases  
$\tilde{B}_a$ to Bob.
\item Alice and Bob determine at which positions the bit strings $\tilde{B}_a$ and  $\tilde{B}_b$ are equal. They discard the bits in 
$\tilde{K}_a$ and $\tilde{K}_b$ where the corresponding bits of $\tilde{B}_a$ and  $\tilde{B}_b$ do not match.
\end{enumerate}
After the execution of the basic BB84 protocol above, the remaining bits of $\tilde{K}_a$ and $\tilde{K}_b$ should be the same, provided that the communication channels used are perfect, and no eavesdropper exists. 

The qMC for the basic BB84 protocol in the simplest case of $n=1$ is depicted in Fig.~\ref{fig:bb84} (right), where  $Set^{|\psi\>}$ is the 1-qubit super-operator which sets the target qubit to $|\psi\>$, $\x=\{X\}$ and $\z=\{Z\}$ are respectively the Pauli-X and Pauli-Z super-operators, and $\a^i = \{|i\>\<i|\}$, $i=0,1, +, -$. As all the super-operators are applied on the same quantum qubit, we omit the subscripts for simplicity. We use the subscripts for the $s$-states to denote the choices of the basis $B_a$ of Alice, the key $K_a$ generated by Alice, and the basis $B_b$ guessed by Bob. For example, in $s_0$, $B_a=0$; in $s_{01}$, $B_a=0$ and $K_a=1$; and in $s_{101}$, $B_a=B_b=1$ and $K_a=0$.
Let $AP=S\cup \{abort\}$ and $L(s)=\{abort\}$ if $s\in \{s_{001}, s_{011}, s_{100}, s_{110}\}$, meaning that at these states Alice and Bob's bases differ, so the protocol will be aborted without generating any key. For other states $s$, we let $L(s)=\{s\}$ naturally.

We use the states \emph{succ} and \emph{fail} to denote the successful and unsuccessful termination of BB84 protocol, respectively. We take the state $s_{101}$ as an example to illustrate the basic idea. As the bases of Alice and Bob are both $\{|+\>, |-\>\}$ at $s_{101}$, they will regard the key bit as the final key generated by the protocol. Thus if the outcome of Bob's measurement is $0$,  which corresponds to the super-operator $\e^+$, then the protocol succeeds since Alice and Bob indeed share the same key bit 0; otherwise the protocol fails as they end up with different bits: Alice with 0 while Bob with 1. That explains why we have $\q(s_{101}, succ) = \e^+$ while $\q(s_{101}, fail) = \e^-$. 
\end{example}
\section{Quantum Computation Tree Logic (QCTL)}

This section is devoted to a quantum extension of 
the probabilistic computation tree
logic (PCTL)~\cite{Hansson:1994kq}, which in turn is an extension of the
classical computation tree logic (CTL)~\cite{DBLP:conf/icalp/MilnerS92}. 

\begin{definition}
The syntax of quantum computation tree logic (QCTL) is as follows:
\begin{eqnarray*}
\Phi &::=& \ a\ |\ \neg \Phi\ |\ \Phi\wedge\Phi\ |\ \Q_{\sim\a}[\phi] \\
\phi&::=& \X\Phi\ |\ \Phi \U^{\leq k} \Phi\ |\ \Phi \U \Phi
\end{eqnarray*}
where $a\in AP$ is an atomic proposition, $\sim {\in}\ \{{\lesssim}, {\gtrsim}\}$, $\a\in \leqI$, and $k\in \N$.
We call
$\Phi$ a \emph{state formula} and $\phi$ a \emph{path
formula}. 
\end{definition}

Compared to the logic presented in~\cite{BCM08}, our QCTL here is simpler and more like PCTL: the only difference is that the formula $\P_{\sim p}[\phi]$ in PCTL, which asserts that the probability of paths from a certain state satisfying the path formula $\phi$ is constrained by $\sim p$ where $0\leq p\leq 1$,  is replaced in QCTL by $\Q_{\sim\a}[\phi]$, which asserts that the accumulated super-operators corresponding to paths from a certain state satisfying the formula $\phi$ is constrained by $\sim \a$ where $0_\h\lesssim \a\lesssim \id_\h$. Note that $\P_{\sim p}[\phi]$ is a special case of $\Q_{\sim\a}[\phi]$ by taking $\e=p\id_\h$.

\begin{definition}\label{def:semantics}
Let $\m=(S, \q,  AP, L)$ be a quantum Markov chain. For any state $s\in S$, the satisfaction relation $\models$ is defined inductively by
\begin{eqnarray*}
s\models a & \mbox{if} & a\in L(s)\\
s\models \neg \Phi&\mbox{if} & s\not\models \Phi\\
s\models \Phi\wedge \Psi &\mbox{if} & s\models \Phi \mbox{ and } s\models \Psi\\
s\models \Q_{\sim \a}[\phi]&\mbox{if} & Q^\m(s,\phi) \sim \a
\end{eqnarray*}
where $$ Q^\m(s,\phi) = Q_s(\{\pi\in \mathit{Path^{\m}(s)} \ |\ \pi\models \phi\}),$$
and for any path $\pi\in \mathit{Path^{\m}(s)}$,
\begin{eqnarray*}
\pi\models \X\Phi & \mbox{   if   } & \pi(1)\models \Phi\\
\pi\models \Phi\U^{\leq k} \Psi&\mbox{ if } & \exists i\in \N.(i\leq k\wedge \pi(i)\models \Psi\wedge \forall j<i.(\pi(j)\models \Phi))\\
\pi\models \Phi\U \Psi&\mbox{ if } & \exists i\in \N.(\pi(i)\models \Psi\wedge \forall j<i.(\pi(j)\models \Phi)).
\end{eqnarray*}
\end{definition}

Similar to PCTL, we can check that for each path formula $\phi$ and each state $s$ in a qMC $\m$, the set $\{\pi\in \mathit{Path^{\m}(s)} \ |\ \pi\models \phi\}$ is in the $\sigma$-algebra generated by $\s^\m(s)$. 
As usual, we introduce some syntactic sugars to simplify notations: the disjunction $\Psi_1\vee \Psi_2 \equiv \neg(\neg\Psi_1\wedge \neg\Psi_2)$, 
the tautology $\true\equiv a\vee \neg a$,
the eventually operator
$\Diamond\Psi\equiv \true\U\Psi$, and the step-bounded eventually operator $\Diamond^{\leq k}\Psi\equiv \true\U^{\leq k}\Psi$.

\begin{example}\label{exm:props}
\rm We revisit the examples in the previous section, to show the expressive power of QCTL.
\begin{enumerate}
\item Example~\ref{exm:qprog}. The QCTL formula
$\Q_{\gtrsim \a}[\Diamond^{\leq k}\ l_3]$
asserts that the probability that the loop program in Example~\ref{exm:qprog} terminates within $k$ iterations is lower bounded by $\e$. That is,
for any initial quantum state $\rho$, the termination probability is not less than $\tr (\a(\rho))$. 
In particular,
the property that it terminates everywhere can be described as
$\Q_{\gtrsim \id_\h}[\Diamond\ l_3].$

\item Example~\ref{exm:bb84}.  The correctness of basic BB84 protocol can be stated as 
$$s\models \Q_{\lesssim 0_\h}[\Diamond \ fail]\wedge \Q_{\gtrsim \frac{1}{2}\id}[\Diamond^{\leq 4}\ succ],$$
which asserts that the protocol never (with probability 0) fails, and with probability at least one half, it will successfully terminate at a shared key within 4 steps. As there is only a half chance for Bob to correctly guess Alice's basis, the probability
of successfully establishing a key cannot exceed 1/2.
\end{enumerate}
\end{example}
\section{Model checking quantum Markov chains}

As in the classical case, given a state $s$ in a qMC $\m=(S, \q,  AP, L)$ and a state formula $\Phi$ expressed in QCTL, model checking if $s$ satisfies $\Phi$ is essentially determining whether $s$ belongs to the satisfaction set $Sat(\Phi)=\{s\in S : s\models \Phi\}$ which is defined inductively as follows:
\begin{eqnarray*}
Sat(a)&=& \{s\in S : a\in L(s)\}\\
Sat(\neg \Psi)&=& S\backslash Sat(\Psi)\\
Sat(\Psi\wedge\Phi)&=& Sat(\Psi)\cap Sat(\Phi)\\
Sat(\Q_{\sim\a} [\phi])&=&\{s\in S : Q^\m(s, \phi) \sim \a\}. 
\end{eqnarray*}
Most of the formulae above are the same as in probabilistic model checking. The only difference is $Sat(\Q_{\sim\a} [\phi])$.  In the following, we will elaborate how to employ the results presented in previous sections to calculate the satisfaction sets for such kind of formulae. To this end, we need to compute $Q^\m(s,\phi)$ for the following three cases\footnote{Strictly speaking, we are computing a super-operator which is $\eqsim$-equivalent to $Q^\m(s,\phi)$. But this is sufficient for our purpose, as only whether or not $Q^\m(s,\phi)\sim \a$ matters here.}.

Case 1: $\phi=\X\Phi$.  By Definition~\ref{def:semantics}, $\{\pi\in \mathit{Path^{\m}(s)} : \pi\models \X\Phi\} = \biguplus_{t\in Sat(\Phi)} Cyl(st)$. Thus 
$$Q^\m(s,\X\Phi)  =Q_s\left(\biguplus_{t\in Sat(\Phi)} Cyl(st)\right)\eqsim \sum_{t\in Sat(\Phi)} Q_s(Cyl(st))= \sum_{t\in Sat(\Phi)} \q(s, t).$$
This can be calculated easily since by the recursive nature of the definition, we can assume that
$Sat(\Phi)$ is already known. 

Case 2: $\phi=\Phi \U^{\leq k}\Psi$. For any $s\in S$ and $k\geq 0$, we let $\Pi^k_s = \{\pi\in \mathit{Path^{\m}(s)} : \pi\models \Phi \U^{\leq k}\Psi\}$.
Then 
$$\Pi_s^k = 
\left\{
\begin{array}{ll}
Cyl(s),  & \mbox{if } s\in Sat(\Psi);  \\
 \emptyset , &\mbox{if } s\not\in Sat(\Phi) \cup Sat(\Psi)\vee (k=0\ \wedge\ s\not\in Sat(\Psi));\\
\biguplus_{t\in post(s)} s^\frown \Pi_t^{k-1}, & \mbox{if  }s\in Sat(\Phi) \backslash Sat(\Psi) \wedge k\geq 1.
\end{array}
\right.
$$
where $post(s)=\{t\in S : \q(s,t)\neq 0_\h\}$, and $s^\frown \Pi_t^{k-1}$ denotes the set of strings obtained by prepending $s$ to strings in $\Pi_t^{k-1}$. By induction on $k$, we can show that for each $k$ and $s$, $\Pi_s^k=\emptyset$ or it is the disjoint union of some cylinder sets; specifically,
we have $\Pi^k_s = \biguplus_{\fpi\in A_s^k} Cyl(\fpi)$ where
$$A_s^k = 
\left\{
\begin{array}{ll}
\{s\},  & \mbox{if } s\in Sat(\Psi);  \\
 \emptyset , &\mbox{if } s\not\in Sat(\Phi) \cup Sat(\Psi)\vee (k=0\ \wedge\ s\not\in Sat(\Psi));\\
\biguplus_{t\in post(s)} s^\frown A_t^{k-1}, & \mbox{if }s\in Sat(\Phi) \backslash Sat(\Psi) \wedge k\geq 1.
\end{array}
\right.
$$
Thus if $s\in Sat(\Phi) \backslash Sat(\Psi)$ and $k\geq 1$, we have
\begin{eqnarray*}
Q_s(\Pi^k_s) &=& Q_s\left(\biguplus_{\fpi\in A_s^k} Cyl(\fpi)\right)\eqsim \sum_{\fpi\in A_s^k} Q_s(Cyl(\fpi))\\
&=&\sum_{t\in post(s)}\sum_{\fpi'\in A_t^{k-1}}   \q(s^\frown \fpi')=\sum_{t\in post(s)}\sum_{\fpi'\in A_t^{k-1}}   \q(\fpi')\q(s,t)\\
&=&\sum_{t\in S}\sum_{\fpi'\in A_t^{k-1}}   Q_t(Cyl(\fpi'))\q(s,t)\eqsim \sum_{t\in S}Q_t(\Pi^{k-1}_t) \q(s,t).
\end{eqnarray*}
Adding the other two cases together, we finally have
$$Q^\m(s, \Phi \U^{\leq k}\Psi) \eqsim 
\left\{
\begin{array}{ll}
\id_\h ,  & \mbox{if } s\in Sat(\Psi);\\
{}0_\h, &\mbox{if }  s\not\in Sat(\Phi) \cup Sat(\Psi)\ \ \vee\\
& \hspace{2em} (k=0\ \wedge\ s\not\in Sat(\Psi));\\
\displaystyle\sum_{t\in S} Q^\m(t, \Phi \U^{\leq k-1}\Psi)\q(s, t), &  \mbox{if $s\in Sat(\Phi) \backslash Sat(\Psi)$}\wedge k\geq 1.
\end{array}
\right.
$$
Again, this can be calculated easily since by the recursive nature of the definition, we can assume that
both $Sat(\Phi)$ and $Sat(\Psi)$ are already known. 

Case 3: $\phi = \Phi \U\Psi$. In this case, we define for any $s\in S$, $\Pi_s = \{\pi\in \mathit{Path^{\m}(s)} : \pi\models \Phi \U\Psi\}$.
Similar to the bounded-until case, we get the equation system
$$Q^\m(s, \Phi \U\Psi) \eqsim  
\left\{
\begin{array}{ll}
\id_\h,  & \mbox{if } s\in Sat(\Psi);  \\
0_\h, &\mbox{if } s\not\in Sat(\Phi) \cup Sat(\Psi);\\
\sum_{t\in S}Q^\m(t, \Phi \U\Psi)\q(s, t) , & \mbox{if } s\in Sat(\Phi) \backslash Sat(\Psi).
\end{array}
\right.
$$
\begin{figure}[t]
\begin{center}
\[
\begin{array}{ll}
\includegraphics[width=0.4\textwidth]{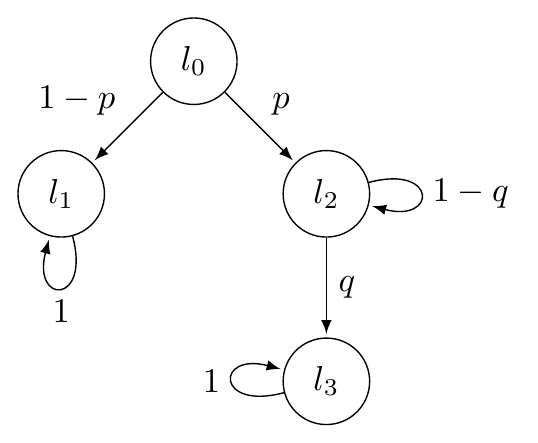}
&\hspace{2em}
\includegraphics[width=0.35\textwidth]{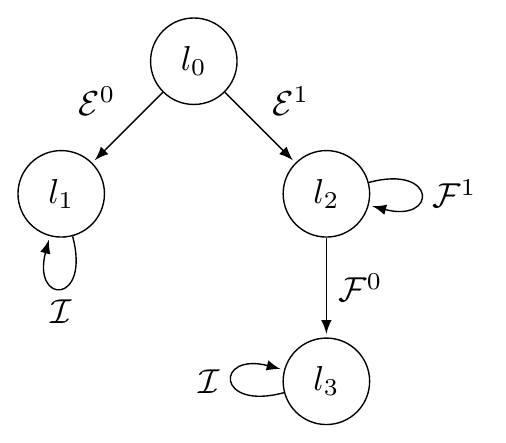}
\end{array}
\]
\end{center}
\caption{Difference between pMC and qMC}\label{fig:pqMC}
\end{figure}

Recall that in probabilistic model checking, to simplify the computation and translate the equation system into one with a unique solution, a pre-computation process is usually employed to compute all the states from which the probability of eventually reaching $Sat(\Psi)$ without leaving states in $Sat(\Phi)$ is exactly 0 or 1. However, it is impossible in quantum case to calculate the \emph{exact} set of states $s$ for which $Q^\m(s, \Phi\U\Psi)$ is equivalent to $0_\h$ or $\id_\h$ without solving an equation system of super-operators. This can be best explained by an example. In Fig.~\ref{fig:pqMC} we have a pMC and a qMC with the same graph structure. Let $Sat(\Psi)=\{l_3\}$ and $Sat(\Phi)=\{l_0, l_2\}$. In the pMC, $P(l_0, \Phi\U\Psi)=0$ if and only if $p=0$ or $q=0$, as $pq=0$ if and only if $p(1-q)^n q=0$ for any $n\geq 0$. However, in the qMC, things are complicated. First, 
we can not claim 
$Q^\m(l_0, \Phi\U\Psi)\not \eqsim 0_\h$ by only checking that neither $\f^0$ nor $\e^1$ is $0_\h$; they can be orthogonal with each other: let $\f^0=\e^0=\{|0\>\<0|\}$,
$\a^1=\f^1=\{|1\>\<1|\}$. Then we have $Q^\m(l_0, \Phi\U\Psi) \eqsim 0_\h$.
Conversely, we can not claim $Q^\m(l_0, \Phi\U\Psi)\eqsim 0_\h$ by only checking $\f^0\a^1=0_\h$ either. Let $\e^0$ and $\e^1$ be defined as above but let 
$\f^0=\{\frac{1}{\sqrt{2}}|0\>\<0|\}$ and $\f^1=\{\frac{1}{\sqrt{2}}|1\>\<1|, \frac{1}{\sqrt{2}}X\}$ where $X$ is the Pauli-X operator. Then $\f^0 \f^1 \a^1\not \eqsim 0_\h$, thus $Q^\m(l_0, \Phi\U\Psi)\not \eqsim 0_\h$. To sum up, to check if  $Q^\m(s, \Phi\U\Psi)\eqsim 0_\h$, the accumulated 
super-operators along all possible paths from $s$ to $Sat(\Psi)$ through $Sat(\Phi)$, including all cycles and self-loops, must be considered. This is essentially as difficult as solving the original super-operator equation system in which the state $s$ is involved. Similar argument applies to determining if $Q^\m(s, \Phi\U\Psi)\eqsim \id_\h$.

We have shown that in general it is not practical to pre-compute the two sets 
$Sat(\Q_{\lesssim 0_\h}[\Phi\U\Psi])$ and 
$Sat(\Q_{\gtrsim \id_\h}[\Phi\U\Psi])$. 
Nevertheless, we can still simplify the calculation by identifying $some$ $S^0$ and $S^\id$ such that
$$S\backslash (Sat(\Psi)\cup Sat(\Phi))\subseteq S^0\subseteq Sat(\Q_{\lesssim 0_\h}[\Phi\U\Psi])$$ 
and $$Sat(\Psi)\subseteq S^\id\subseteq Sat(\Q_{\gtrsim \id_\h}[\Phi\U\Psi]),$$ which are calculated by the algorithms presented in
Table~{\ref{tb:s01}, motivated by~\cite{KNP07}. 
\newsavebox{\tablebox}
\begin{table}[t]
\begin{lrbox}{\tablebox}
\begin{tabular}[t]{l}
\begin{tabular}{l|l}
\hline\\
\begin{tabular}{l}
Input: $Sat(\Phi)$ and $Sat(\Psi)$\\
Output: A subset $S^0$ of $S$ such that\\
\hspace{4em}$S\backslash Sat(\Psi)\backslash Sat(\Phi)\subseteq S^0\subseteq Sat(\Q_{\lesssim 0_\h}[\Phi\U\Psi]) $ \\
\\
\hline\\
$R:=\{s : \mbox{no direct path from $s$ to states in }Sat(\Psi)\}$\\
$R:=R\cup (S\backslash Sat(\Phi)\backslash Sat(\Psi))$\\
$done := \bold{false}$\\
$\while\ (done = \bold{false})\ \ddo$\\
\hspace{1em}$R':=R\cup \{s \in S\backslash R: \sum_{t\in R}\q(s,t) + \q(s,s)\eqsim \id\}$\\
\hspace{1em}$\iif\ (R'=R)\ \then\ done := \bold{true}$\\
\hspace{1em}$R:=R'$\\
$\bold{od}$\\
return $R$
\end{tabular}
&\begin{tabular}{l}
Input: $Sat(\Phi)$ and $Sat(\Psi)$\\
Output: A subset $S^\id$ of $S$ such that\\
\hspace{4em}$Sat(\Psi)\subseteq S^\id\subseteq Sat(\Q_{\gtrsim \id_\h}[\Phi\U\Psi]) $ \\
\\
\hline\\
$R:=Sat(\Psi)$\\
$done := \bold{false}$\\
$\while\ (done = \bold{false})\ \ddo$\\
\hspace{1em}$R':=R\cup \{s \in Sat(\Phi)\backslash R: \sum_{t\in R}\q(s,t) \eqsim \id\}$\\
\hspace{1em}$\iif\ (R'=R)\ \then\ done := \bold{true}$\\
\hspace{1em}$R:=R'$\\
$\bold{od}$\\
return $R$\\
\\
\end{tabular}\\
\\
\hline
\end{tabular}
\end{tabular}
\end{lrbox}
\resizebox{\textwidth}{!}{\usebox{\tablebox}}\\
\caption{Algorithms to calculate $S^0$ and $S^\id$.}\label{tb:s01}
\end{table}%
Now let $S^?=S\backslash (S^0\cup S^\id).$
Then for each $s\in S^?$, the argument for the bounded-until case indeed shows for $k\geq 0$,
\begin{equation}\label{eq:tmp01}
Q^\m(s, \Phi \U^{\leq k+1}\Psi) \lesssim  \sum_{t\in S^?} Q^\m(t, \Phi \U^{\leq k}\Psi)\q(s, t) + \sum_{t\in S^\id} \q(s,t)
\end{equation}
and for unbounded-until case,
\begin{equation}\label{eq:tmp02}
Q^\m(s, \Phi \U\Psi) \eqsim  \sum_{t\in S^?} Q^\m(t, \Phi \U\Psi)\q(s, t) + \sum_{t\in S^\id} \q(s,t)
\end{equation}
Let $\t = (\q(t,s))_{s, t\in S^?}$ and $\widetilde{\g} = (\sum_{t\in S^\id} \q(s,t))_{s\in S^?}$ be two state-indexed matrices of super-operators. 
\begin{theorem}\label{thm:5.1} Let $
f(X)=X\t + \widetilde{\g}
$, and $\widetilde{\e}$ be the least fixed point of $f$. Then for 
 any $s\in S^?$, \[\widetilde{\e}_s \eqsim Q^\m(s, \Phi \U\Psi).\]
\end{theorem}
{\it Proof.}
First, we check that
for each $s\in S^?$, $$\sum_{t\in S^?} \t_{t,s} + \widetilde{\g}_s =\sum_{t\in S^?} \q(s,t) + \sum_{t\in S^\id} \q(s,t)=\sum_{t\in S^\id\uplus S^?} \q(s,t) \lesssim \id_\h.$$
The existence of the least fixed point $\widetilde{\e}$ in $\leqI^{|S^?|}$ follows from Theorem~\ref{thm:fixpoint}. 
Let $\widetilde{\e}^{(0)} = (\widetilde{\e}^{(0)}_s)_{s\in S^?}$ where $\widetilde{\e}^{(0)}_s = 0_\h$ for each $s\in S^?$, and $\widetilde{\e}^{(k+1)} = \widetilde{\e}^{(k)}\t + \widetilde{\g}$. As $f$ is Scott continuous with respect to $\le$, we have $ \widetilde{\e}^{(0)}\le  \widetilde{\e}^{(1)}\le \dots$, and $\widetilde{\e} = \lim_{k} \widetilde{\e}^{(k)}$. On the other hand, by Lemma~\ref{lem:distprop}(4), for any $s\in S^?$ there exists a nondecreasing sequence $(\f^k_s)_{k\geq 0}$ of super-operators such that $Q^\m(s, \Phi \U^{\leq k}\Psi)\eqsim\f^k_s$ and $Q^\m(s, \Phi \U\Psi)\eqsim \lim_k \f^k_s$. Thus to prove the theorem, it suffices to show that for any $k\geq 0$ and $s\in S^?$,
\begin{equation}\label{eq:ind}
Q^\m(s, \Phi \U^{\leq k}\Psi)\lesssim \widetilde{\e}_s^{(k)} \lesssim Q^\m(s, \Phi \U\Psi).
\end{equation}

In the following, we prove Eq.(\ref{eq:ind}) by induction. Fix arbitrarily $s\in S^?$. When $k=0$, we have  $\widetilde{\e}_s^{(0)}  =Q^\m(s, \Phi \U^{\leq 0}\Psi)= 0_\h\lesssim Q^\m(s, \Phi \U\Psi)$ as $s\not\in Sat(\Psi)$. Now suppose Eq.(\ref{eq:ind}) holds for $k$. Then 
\begin{eqnarray*}
Q^\m(s, \Phi \U^{\leq k+1}\Psi) &\lesssim &  \sum_{t\in S^?} Q^\m(t, \Phi \U^{\leq k}\Psi) \q(s, t) + \sum_{t\in S^\id} \q(s,t).  \hspace{9.1em}  \mbox{by Eq.(\ref{eq:tmp01})}\\
&\lesssim & \sum_{t\in S^?} \widetilde{\e}_t^{(k)} \q(s, t) + \sum_{t\in S^\id} \q(s,t) \hspace{6em}  \mbox{by induction and Lemma~\ref{lem:rightapp}}\\
&= &\widetilde{\e}_s^{(k+1)}
\end{eqnarray*}
and
\begin{eqnarray*}
\widetilde{\e}_s^{(k+1)} &=& \sum_{t\in S^?} \widetilde{\e}_t^{(k)} \q(s, t) + \sum_{t\in S^\id} \q(s,t)\\
&\lesssim & \sum_{t\in S^?} Q^\m(t, \Phi \U\Psi) \q(s, t) + \sum_{t\in S^\id} \q(s,t) \hspace{6em}  \mbox{by induction and Lemma~\ref{lem:rightapp}}\\
&\eqsim & Q^\m(s, \Phi \U\Psi).  \hspace{25.1em}  \mbox{by Eq.(\ref{eq:tmp02})}
\end{eqnarray*}
\qed

From Theorems~\ref{thm:5.1} and \ref{thm:fixpoint}, whether or not 
$Q^\m(s, \Phi \U\Psi)\sim \a$ can be determined efficiently.

\begin{example}
\rm This example is devoted to model checking the properties listed in Example~\ref{exm:props} against the qMCs of Examples~\ref{exm:qprog} and \ref{exm:bb84}. 
\begin{enumerate}
\item Quantum loop program. We only check the property $\Q_{\gtrsim \e}[\Diamond\ l_3]$. Let $\f=\{|+\>\<i| : i=0,1\}$ be the super-operator which sets the target qubit to $|+\>=(|0\>+|1\>)/\sqrt{2}$, $\e^i = \{|i\>\<i|\}$, $i=0,1$, and $\e = \x$ the Pauli-X super-operator. 
We first calculate that $Sat(l_3)=\{l_3\}$ and $Sat(\true)=\{l_0, l_1, l_2, l_3\}$. Then from the algorithm in Table~\ref{tb:s01} we have $S^0 =\emptyset$, and $S^\id=\{l_3\}$. So $S^?=\{l_0, l_1, l_2\}$.
We proceed as follows:
\begin{eqnarray*}
Q^{\m}(l_0, \Diamond\ l_3) &=& Q^{\m}(l_1, \Diamond\ l_3) \f\\
Q^{\m}(l_1, \Diamond\ l_3) &=& Q^{\m}(l_2, \Diamond\ l_3) \e^1 + \e^0\\
Q^{\m}(l_2, \Diamond\ l_3) &=& Q^{\m}(l_1, \Diamond\ l_3) \e
\end{eqnarray*}
Let $\widetilde{\g} = [0_\h, \e^0, 0_\h]$ with its matrix representation being $M_{\widetilde{\g}}= [0_{4\times 4}, M_{\e^0}, 0_{4\times 4}]$, and 
$$\t = 
\left(
\begin{array}{cccc}
  0_\h   &  0_\h &  0_\h  \\
 \f &  0_\h  &  \e \\
  0_\h   & \e^1 &  0_\h 
\end{array}
\right)
\mbox{ with } M_\t = 
\left(
\begin{array}{cccc}
  0_{4\times 4}   &  0_{4\times 4} & 0_{4\times 4}  \\
 M_{\f} &  0_{4\times 4}  &   M_{\e} \\
  0_{4\times 4}   & M_{\e^1} &  0_{4\times 4} 
\end{array}
\right),
$$
where
$$M_{\e^0}=|0\>\<0|\otimes |0\>\<0|=\left(
\begin{array}{cccc}
  1   &  0 & 0 & 0   \\
  0  &  0 & 0 & 0   \\
  0  &  0 & 0 & 0   \\
  0  &  0 & 0 & 0   
\end{array}
\right),\ M_{\e^1}=|1\>\<1|\otimes |1\>\<1|=\left(
\begin{array}{cccc}
  0   &  0 & 0 & 0   \\
  0  &  0 & 0 & 0   \\
  0  &  0 & 0 & 0   \\
  0  &  0 & 0 & 1   
\end{array}
\right),
$$
$$M_{\e}=X\otimes X=\left(
\begin{array}{cccc}
  0   &  0 & 0 & 1   \\
  0  &  0 & 1 & 0   \\
  0  &  1 & 0 & 0   \\
  1  &  0 & 0 & 0   
\end{array}
\right),\ M_{\f}=\sum_{i=0}^1|+\>\<i| \otimes |+\>\<i|=\frac{1}{2}\left(
\begin{array}{cccc}
  1   &  0 & 0 & 1   \\
  1  &  0 & 0 & 1   \\
  1  &  0 & 0 & 1   \\
  1  &  0 & 0 & 1   
\end{array}
\right).
$$
Using Matlab, we find that all eigenvalues of the matrix $M_\t$ have the absolute value strictly less than 1, and
$M_{\widetilde{\g}}(I-M_\t)^{-1} = [M, M, M]$ where 
$$M=\left(
\begin{array}{cccc}
  1   &  0 & 0 & 1   \\
  0  &  0 & 0 & 0   \\
  0  &  0 & 0 & 0   \\
  0  &  0 & 0 & 0   
\end{array}
\right)=|0\>\<0|\otimes |0\>\<0| + |0\>\<1|\otimes |0\>\<1|.
$$
Thus for $i=0,1,2$, $Q^{\m}(l_i, \Diamond\ l_3) = Set^0$ where
$Set^0
= \{|0\>\<0|, |0\>\<1|\}\eqsim \id$, and so
$l_i\models \Q_{\gtrsim \a}[\Diamond\ l_3]$ for any $\a\lesssim \id$.

\item BB84 protocol. We will compute $Q^{\m}(s, \Diamond^{\leq 4}\ succ)$ and  $Q^{\m}(s, \Diamond\ fail)$ separately. For $Q^{\m}(s, \Diamond^{\leq 4}\ succ)$, we first obtain from the algorithm in Table~\ref{tb:s01} that 
$S^0 =\{s_{001}, s_{011}, s_{100}, s_{110}, fail\}$, and $S^\id=\{succ\}$. Then
Table~\ref{tb:qmt} calculates $Q^{\m}(t, \Diamond^{\leq k}\ succ)$ for each $t\in S^?=S\backslash S^0\backslash S^\id$ and $0\leq k\leq 4$. The item $Q^{\m}(s, \Diamond^{\leq 4}\ succ)$ is calculated as follows:
\begin{eqnarray*}
Q^{\m}(s, \Diamond^{\leq 4}\ succ) &=& \sum_{t\in S^?}Q^{\m}(t, \Diamond^{\leq 3}\ succ)\q(s, t)\\
&=&
\frac{1}{8}(\e^0 +\e^1\x)Set^{|0\>} + \frac{1}{8}(\e^+ +\e^-\z)Set^{|+\>} \\
&=& \frac{1}{8}(Set^{|0\>} + Set^{|1\>}+ Set^{|+\>} + Set^{|-\>})\eqsim \frac{1}{2}\id_\h. 
\end{eqnarray*}

Similarly, for $Q^{\m}(t, \Diamond\ fail)$ we have
$S^0 =\{s_{001}, s_{011}, s_{100}, s_{110}, succ\}$ and $S^\id=\{fail\}$. Table~\ref{tb:qmtf} computes $Q^{\m}(t, \Diamond^{\leq 4}\ fail)$ for each $t\in S^?=S\backslash S^0\backslash S^\id$ and $0\leq k\leq 4$ where
\begin{eqnarray*}
Q^{\m}(s, \Diamond^{\leq 4}\ fail) &=& \sum_{t\in S}Q^{\m}(t, \Diamond^{\leq 3}\ fail)\q(s, t)\\
&=&
\frac{1}{8}(\e^1 +\e^0\x)Set^{|0\>} + \frac{1}{8}(\e^- +\e^+\z)Set^{|+\>} = 0_\h. 
\end{eqnarray*}
Note that $Q^{\m}(t, \Diamond^{\leq k}\ fail)=Q^{\m}(t, \Diamond^{\leq 4}\ fail)$ for any $t\in S$ and $k>4$. We have  $$s\models \Q_{\lesssim 0_\h}[\Diamond \ fail]\wedge \Q_{\gtrsim \frac{1}{2}\id}[\Diamond^{\leq 4}\ succ]$$ as expected.

\end{enumerate}
\end{example}

\renewcommand{\arraystretch}{1.5}
\begin{table}[t]
\begin{lrbox}{\tablebox}
\begin{tabular}{c||*{11}{c|}c}
\hline
\backslashbox{$k$}{$t$}
&$s$ & $s_{0}$& $s_{1}$& $s_{00}$& $s_{01}$& $s_{10}$ & $s_{11}$& $s_{000}$& $s_{010}$& $s_{101}$& $s_{111}$ & $succ$
\\
\hline\hline
$0$ & $0_\h$& $0_\h$& $0_\h$& $0_\h$& $0_\h$ & $0_\h$&$0_\h$& $0_\h$& $0_\h$& $0_\h$& $0_\h$& $\id_\h$
\\
\hline
$1$ & $0_\h$& $0_\h$& $0_\h$& $0_\h$& $0_\h$ & $0_\h$&$0_\h$& $\e^0$& $\e^1$& $\e^+$& $\e^-$& $\id_\h$
\\
\hline
$2$ & $0_\h$& $0_\h$& $0_\h$& $\frac{1}{2}\e^0$& $\frac{1}{2}\e^1$ & $\frac{1}{2}\e^+$&$\frac{1}{2}\e^-$& $\e^0$& $\e^1$& $\e^+$& $\e^-$& $\id_\h$
\\
\hline
$3$ & $0_\h$& $\frac{1}{4}(\e^0 +\e^1\x)$& $\frac{1}{4}(\e^+ + \e^-\z)$& $\frac{1}{2}\e^0$& $\frac{1}{2}\e^1$ & $\frac{1}{2}\e^+$&$\frac{1}{2}\e^-$& $\e^0$& $\e^1$& $\e^+$& $\e^-$& $\id_\h$
\\
\hline
$4$ & $\displaystyle\frac{1}{8}\sum_{i\in\{0, 1, +, -\}} Set^{|i\>}$& $\frac{1}{4}(\e^0 + \e^1\x)$& $\frac{1}{4}(\e^+ + \e^-\z)$& $\frac{1}{2}\e^0$& $\frac{1}{2}\e^1$ & $\frac{1}{2}\e^+$&$\frac{1}{2}\e^-$& $\e^0$& $\e^1$& $\e^+$& $\e^-$& $\id_\h$
\\
\hline
\end{tabular}
\end{lrbox}
\resizebox{\textwidth}{!}{\usebox{\tablebox}}\\
\caption{Table for $Q^{\m}(t, \Diamond^{\leq k}\ succ), 0\leq k\leq 4$.}\label{tb:qmt}
\end{table}%

\begin{table}[t]
\begin{lrbox}{\tablebox}
\begin{tabular}{c||*{10}{c|}c}
\hline
\backslashbox{$k$}{$t$}&$\  s\ $ & $s_{0}$& $s_{1}$& $\ s_{00}\ $&\ $s_{01}$\ & $s_{10}$ & $s_{11}$& $s_{000}$& $s_{010}$& $s_{101}$& $s_{111}$ 
\\
\hline\hline
$0$ & $0_\h$& $0_\h$& $0_\h$& $0_\h$& $0_\h$ & $0_\h$&$0_\h$& $0_\h$& $0_\h$& $0_\h$& $0_\h$
\\
\hline
$1$ & $0_\h$& $0_\h$& $0_\h$& $0_\h$& $0_\h$ & $0_\h$&$0_\h$& $\e^1$& $\e^0$& $\e^-$& $\e^+$
\\
\hline
$2$ & $0_\h$& $0_\h$& $0_\h$& $\frac{1}{2}\e^1$& $\frac{1}{2}\e^0$ & $\frac{1}{2}\e^-$&$\frac{1}{2}\e^+$& $\e^1$& $\e^0$& $\e^-$& $\e^+$
\\
\hline
$3$ & $0_\h$& $\frac{1}{4}(\e^1 +\e^0\x)$& $\frac{1}{4}(\e^- + \e^+\z)$& $\frac{1}{2}\e^1$& $\frac{1}{2}\e^0$ & $\frac{1}{2}\e^-$&$\frac{1}{2}\e^+$& $\e^1$& $\e^0$& $\e^-$& $\e^+$
\\
\hline
$4$ & $0_\h$& $\frac{1}{4}(\e^1 +\e^0\x)$& $\frac{1}{4}(\e^- + \e^+\z)$& $\frac{1}{2}\e^1$& $\frac{1}{2}\e^0$ & $\frac{1}{2}\e^-$&$\frac{1}{2}\e^+$& $\e^1$& $\e^0$& $\e^-$& $\e^+$
\\
\hline
\end{tabular}
\end{lrbox}
\resizebox{\textwidth}{!}{\usebox{\tablebox}}\\
\caption{Table for $Q^{\m}(t, \Diamond^{\leq k}\ fail), 0\leq k\leq 4$.}\label{tb:qmtf}
\end{table}%

\subsection{Complexity}

Recall that the overall time complexity for model checking a PCTL formula $\Phi$ against a
classical Markov chain with $n$ states is linear in $|\Phi|$ and polynomial in $n$, where the size $|\Phi|$ of
a formula is defined to be the number of logical connectives
and temporal operators in the formula plus the sum of the sizes of the temporal
operators~\cite{KNP07}. Let $d=dim(\h)$. The greatest extra cost of our algorithm is the until operator but from Theorem~\ref{thm:fixpoint} it is of the order $n^2d^4$. Thus the complexity for model checking a QCTL formula $\Phi$ against a qMC is again linear in $|\Phi|$ and polynomial in $n$ and $d$.

\section{Conclusion and future work}

The main contribution of this paper is a novel notion of quantum Markov chains where
quantum effects are entirely encoded into super-operators labelling transitions, while leaving the
location information being classical. By employing Kluvanek's generalisation of Carath\'eodory-Hahn extension theorem from vector measure theory,
we are able to define an appropriate super-operator valued measure on events of infinite paths in a quantum Markov chain.
Based on this, we propose a quantum extension of PCTL and develop an
algorithm for model-checking quantum Markov chains.

We demonstrate the utility of the techniques developed in this paper by examples of model-checking the correctness of a simple quantum loop program as well as the basic BB84 protocol.
The properties checked in these examples are essentially classical in the sense that we are interested only in the \emph{probabilities} of certain events, not the quantum states themselves.
However, there are also quantum protocols, such as \emph{teleportation} which can make use of a maximally
entangled state shared between the sender and the receiver to teleport an unknown
quantum state by sending only classical information~\cite{BB93}, where the properties we need to check are related to the resultant quantum states directly. 
One possible way of verifying such properties is to extend the atomic propositions, say, in the case of teleportation, to specify whether or not the accumulated super-operator from the initial location to the current location is proportional to the identity super-operator. This treatment resorts to  
the ability to calculate the accumulated super-operators when constructing the model, but this can be done by the model-checking algorithm presented in this paper. 
We will further explore the possibility of this idea in our future work. 

The BB84 protocol is verified in this paper with the implicit assumption that no noise occurs at all. As pointed out in the Introduction, security of quantum cryptography is always compromised by
the inevitable noise in physical implementations. A natural question then arises: can the techniques in the current paper be used to check security of physically implemented quantum cryptographic systems? This is also one of the future directions we are pursuing. It seems that the quantum Markov chain proposed in this paper is inclusive enough for this purpose because noisy implementation of unitary operators used in quantum communication can always be modelled by super-operators.

\section*{Acknowledgement}
This work was supported by Australian Research Council grants DP110103473, DP130102764, and FT100100218.
The authors are also partially supported by the Overseas Team Program of Academy of Mathematics and Systems Science, Chinese Academy of Sciences.

\bibliographystyle{plain}

\newpage
\section{Appendix}

\subsection{Basic linear algebra}
A {\it Hilbert space} $\h$ is a complete vector space equipped with an inner
product $$\langle\cdot|\cdot\rangle:\h\times \h\rightarrow \mathbf{C}$$
such that 
\begin{enumerate}
\item
$\langle\psi|\psi\rangle\geq 0$ for any $|\psi\>\in\h$, with
equality if and only if $|\psi\rangle =0$;
\item
$\langle\phi|\psi\rangle=\langle\psi|\phi\rangle^{\ast}$;
\item
$\langle\phi|\sum_i c_i|\psi_i\rangle=
\sum_i c_i\langle\phi|\psi_i\rangle$,
\end{enumerate}
where $\mathbf{C}$ is the set of complex numbers, and for each
$c\in \mathbf{C}$, $c^{\ast}$ stands for the complex
conjugate of $c$. For any vector $|\psi\rangle\in\h$, its
length $|||\psi\rangle||$ is defined to be
$\sqrt{\langle\psi|\psi\rangle}$, and it is said to be {\it normalised} if
$|||\psi\rangle||=1$. Two vectors $|\psi\>$ and $|\phi\>$ are
{\it orthogonal} if $\<\psi|\phi\>=0$. An {\it orthonormal basis} of a Hilbert
space $\h$ is a basis $\{|i\rangle\}$ where each $|i\>$ is
normalised and any pair of them are orthogonal.

Let $\lh$ be the set of linear operators on $\h$.  For any $A\in
\lh$, $A$ is {\it Hermitian} if $A^\dag=A$ where
$A^\dag$ is the adjoint operator of $A$ such that
$\<\psi|A^\dag|\phi\>=\<\phi|A|\psi\>^*$ for any
$|\psi\>,|\phi\>\in\h$. The fundamental {\it spectral theorem} states that
a set of normalised eigenvectors of a Hermitian operator in
$\lh$ constitutes an orthonormal basis for $\h$. That is, there exists
a so-called spectral decomposition for each Hermitian $A$ such that
$$A=\sum_i\lambda_i |i\>\<i|=\sum_{\lambda_{i}\in spec(A)}\lambda_i E_i$$
where the set $\{|i\>\}$ constitute an orthonormal basis of $\h$, $spec(A)$ denotes the set of
eigenvalues of $A$,
and $E_i$ is the projector to
the corresponding eigenspace of $\lambda_i$.
A linear operator $A\in \lh$ is {\it unitary} if $A^\dag A=A A^\dag=I_\h$ where $I_\h$ is the
identity operator on $\h$. 
The {\it  trace} of $A$ is defined as $\tr(A)=\sum_i \<i|A|i\>$ for some
given orthonormal basis $\{|i\>\}$ of $\h$. It is worth noting that
trace function is actually independent of the orthonormal basis
selected. It is also easy to check that trace function is linear and
$\tr(AB)=\tr(BA)$ for any operators $A,B\in \lh$.

Let $\h_1$ and $\h_2$ be two Hilbert spaces. Their {\it tensor product} $\h_1\otimes \h_2$ is
defined as a vector space consisting of
linear combinations of the vectors
$|\psi_1\psi_2\rangle=|\psi_1\>|\psi_2\rangle =|\psi_1\>\otimes
|\psi_2\>$ with $|\psi_1\rangle\in \h_1$ and $|\psi_2\rangle\in
\h_2$. Here the tensor product of two vectors is defined by a new
vector such that
$$\left(\sum_i \lambda_i |\psi_i\>\right)\otimes
\left(\sum_j\mu_j|\phi_j\>\right)=\sum_{i,j} \lambda_i\mu_j
|\psi_i\>\otimes |\phi_j\>.$$ Then $\h_1\otimes \h_2$ is also a
Hilbert space where the inner product is defined as the following:
for any $|\psi_1\>,|\phi_1\>\in\h_1$ and $|\psi_2\>,|\phi_2\>\in
\h_2$,
$$\<\psi_1\otimes \psi_2|\phi_1\otimes\phi_2\>=\<\psi_1|\phi_1\>_{\h_1}\<
\psi_2|\phi_2\>_{\h_2}$$ where $\<\cdot|\cdot\>_{\h_i}$ is the inner
product of $\h_i$. For any $A_1\in \mathcal{L}(\h_1)$ and $A_2\in
\mathcal{L}(\h_2)$, $A_1\otimes A_2$ is defined as a linear operator
in $\mathcal{L}(\h_1 \otimes \h_2)$ such that for each
$|\psi_1\rangle \in \h_1$ and $|\psi_2\rangle \in \h_2$,
$$(A_1\otimes A_2)|\psi_1\psi_2\rangle = A_1|\psi_1\rangle\otimes
A_2|\psi_2\rangle.$$  The {\it partial trace} of $A\in\mathcal{L}(\h_1
\otimes \h_2)$ with respected to $\h_1$ is defined as
$\tr_{\h_1}(A)=\sum_i \<i|A|i\>$ where $\{|i\>\}$ is an orthonormal
basis of $\h_1$. Similarly, we can define the partial trace of $A$
with respected to $\h_2$. Partial trace functions are also
independent of the orthonormal basis selected.

Traditionally, a linear operator $\e$ on $\lh$ is called a $super$-$operator$ on $\h$.
A super-operator is said to be 
{\it completely positive} if it maps
positive operators in $\mathcal{L}(\h)$ to positive operators in
$\mathcal{L}(\h)$, and for any auxiliary Hilbert space $\h'$, the
trivially extended operator $\mathcal{I}_{\h'}\otimes \e$ also maps
positive operators in $\mathcal{L(H'\otimes H)}$ to positive
operators in $\mathcal{L(H'\otimes H)}$. Here $\mathcal{I}_{\h'}$ is
the identity operator on $\mathcal{L(H')}$. We always assume complete positivity for super-operators in this paper. The elegant and powerful
{\it Kraus representation theorem} \cite{Kr83} of completely positive
super-operators states that a super-operator $\e$ is completely positive
if and only if there are some set of operators $\{E_i : i\in I\}$ with appropriate dimension such that
$$
\e(A)=\sum_{i\in I} E_iA E_i^\dag
$$
for any $A\in \lh$. The operators $E_i$ are called Kraus operators
of $\e$. We abuse the notation slightly by denoting $\e=\{E_i : i\in I\}$. It is worth noting that the set of
Kraus operators is not unique and we can always take one such that the number of Kraus operators does not 
exceed $d^2$ where $d$ is the dimension of the Hilbert space. 
A super-operator is said to be
{\it trace-preserving} if $\tr(\e(A))= \tr(A)$ for any positive $A\in \lh$; equivalently, its Kraus operators $E_i$ satisfy $\sum_i E_i^\dag E_i= I$.
In  this paper, we will
use the well-known (unitary) Pauli super-operators
$\x=\{X\},\z=\{Z\}$, and $\y=\{Y\}$ where
\[
X=\left(%
\begin{array}{cc}
  0 & 1 \\
  1 & 0 \\
\end{array}%
\right),\ Z=\left(%
\begin{array}{cc}
  1 & 0 \\
  0 & -1 \\
\end{array}%
\right),\ Y=\left(%
\begin{array}{cc}
  0 & -i \\
  i & 0 \\
\end{array}%
\right).
\]
\subsection{Basic quantum mechanics}

According to von Neumann's formalism of quantum mechanics
\cite{vN55}, an isolated physical system is associated with a
Hilbert space which is called the {\it state space} of the system. A {\it pure state} of a
quantum system is a normalised vector in its state space, and a
{\it mixed state} is represented by a density operator on the state
space. Here a density operator $\rho$ on Hilbert space $\h$ is a
positive linear operator such that $\tr(\rho)= 1$. 
Another
equivalent representation of density operator is probabilistic
ensemble of pure states. In particular, given an ensemble
$\{(p_i,|\psi_i\rangle)\}$ where $p_i \geq 0$, $\sum_{i}p_i=1$,
and $|\psi_i\rangle$ are pure states, then
$\rho=\sum_{i}p_i|\psi_i\>\langle\psi_i|$ is a density
operator.  Conversely, each density operator can be generated by an
ensemble of pure states in this way.  Let $\dh$ denote the set of
density operators on $\h$.

The state space of a composite system (for example, a quantum system
consisting of many qubits) is the tensor product of the state spaces
of its components. For a mixed state $\rho$ on $\h_1 \otimes \h_2$,
partial traces of $\rho$ have explicit physical meanings: the
density operators $\tr_{\h_1}\rho$ and $\tr_{\h_2}\rho$ are exactly
the reduced quantum states of $\rho$ on the second and the first
component system, respectively. Note that in general, the state of a
composite system cannot be decomposed into tensor product of the
reduced states on its component systems. A well-known example is the
 2-qubit state
$$|\Psi\>=\frac{1}{\sqrt{2}}(|00\>+|11\>).
$$
This kind of state is called {\it entangled state}.
To see the strangeness of entanglement, suppose a measurement $M=
\lambda_0[|0\>]+\lambda_1[|1\>]$ is applied on the first qubit
of $|\Psi\>$ (see the following for the definition of
quantum measurements). Then after the measurement, the second qubit will
definitely collapse into state $|0\>$ or $|1\>$ depending on whether
the outcome $\lambda_0$ or $\lambda_1$ is observed. In other words,
the measurement on the first qubit changes the state of the second
qubit in some way. This is an outstanding feature of quantum mechanics
which has no counterpart in classical world, and is the key to many
quantum information processing tasks  such as teleportation
\cite{BB93} and superdense coding \cite{BW92}.

The evolution of a closed quantum system is described by a unitary
operator on its state space: if the states of the system at times
$t_1$ and $t_2$ are $\rho_1$ and $\rho_2$, respectively, then
$\rho_2=U\rho_1U^{\dag}$ for some unitary operator $U$ which
depends only on $t_1$ and $t_2$. In contrast, the general dynamics which can occur in a physical system is
described by a trace-preserving super-operator on its state space. 
Note that the unitary transformation $U(\rho)=U\rho U^\dag$ is
a super-operator. 

A quantum {\it measurement} is described by a
collection $\{M_m\}$ of measurement operators, where the indices
$m$ refer to the measurement outcomes. It is required that the
measurement operators satisfy the completeness equation
$\sum_{m}M_m^{\dag}M_m=I_\h$. If the system is in state $\rho$, then the probability
that measurement result $m$ occurs is given by
$$p(m)=\tr(M_m^{\dag}M_m\rho),$$ and the state of the post-measurement system
is $M_m\rho M_m^{\dag}/p(m).$ 

A particular case of measurement is {\it projective measurement} which is usually represented by a Hermitian operator.  Let  $M$ be a
Hermitian operator and
\begin{equation}\label{eq:specdec}
M=\sum_{m\in spec(M)}mE_m
\end{equation} 
its spectral decomposition. Obviously, the projectors  $\{E_m:m\in
spec(M)\}$ form a quantum measurement. If the state of a quantum
system is $\rho$, then the probability that result $m$ occurs when
measuring $M$ on the system is $p(m)=\tr(E_m\rho),$ and the
post-measurement state of the system is $E_m\rho E_m/p(m).$
Note that for each outcome $m$, the map $$\e_m(\rho) =
E_m\rho E_m$$
is again a super-operator by Kraus Theorem.

\subsection{Proof of Theorem~\ref{thm:fixpoint}}\label{app:2.5}

This section is devoted to the proof of Theorem~\ref{thm:fixpoint}. We first recall some basic results from linear algebra.
Let $M$ be a squared matrix and $M = S J S^{-1}$ its Jordan decomposition where $S$ is a
nonsingular matrix,  
$$J = diag (J_{n_1}(\lambda_1), J_{n_2}(\lambda_2), \dots, J_{n_k}(\lambda_k)),$$
and each $J_{n_i}(\lambda_i)$ is an $n_i\times n_i$-Jordan block with the eigenvalue $\lambda_i$. Let $\tilde{M}= S \tilde{J} S^{-1}$ where $\tilde{J}$ is obtained from $J$ by replacing each Jordan block whose associated eigenvalue has absolute value greater than or equal to 1 with the zero block of the same size; that is $\tilde{J} = diag (J^1, J^2, \dots, J^k)$ where
$$J^i = 
\left\{
\begin{array}{ll}
  0_{n_i\times n_i},  & \mbox{if } |\lambda_i|\geq 1;  \\
  \\
 J_{n_i}(\lambda_i), & \mbox{otherwise}.
\end{array}
\right.
$$

\begin{lemma}\label{lem:limit}
If $\sum_{m=0}^{\infty} NM^m$ exists, then for each $m\geq 0$, $NM^m = N\tilde{M}^m$.
\end{lemma}
{\it Proof.}
Suppose $\sum_{m=0}^{\infty} NM^m$ exists. Then $\lim_{m\ra \infty} NM^m=0$. Since $M^m = SJ^m S^{-1}$,  we have $\lim_{m\ra \infty} NSJ^m=0$.
Decompose the columns of $NS$ according to the blocks of $J$ as
$NS = (K_{1}, \dots, K_k)$. Then 
$$NSJ^m = 
\left(K_{1}J^m_{n_1}(\lambda_1),    \cdots,  K_{k}J^m_{n_k}(\lambda_k)   
\right),
$$
and for each $i$, $\lim_{m\ra \infty} K_{i}J^m_{n_i}(\lambda_i)=0$. Let $\|\cdot \|$ be an (arbitrary) matrix norm. Then
\begin{eqnarray*}
0&=&\lim_{m\ra \infty} \| K_{i}J^m_{n_i}(\lambda_i)\|\\
& = &\lim_{m\ra \infty}\| K_{i}\|\cdot \|J_{n_i}(\lambda_i)\|^m\\
&\geq&\lim_{m\ra \infty}\| K_{i}\|\cdot |\lambda_i|^m
\end{eqnarray*} where the last inequality is from Theorem 5.6.9 of~\cite{HJ90}. 
From this we deduce that $K_{i}=0$ for each $i$ whenever $ |\lambda_i|\geq 1$. Thus $NSJ^m=NS\tilde{J}^m$, and $NM^m = N\tilde{M}^m$.
\hfill $\Box$

\begin{corollary}\label{cor:limit} For any matrices $N$ and $M$,
$$\sum_{m=0}^{\infty} NM^m=N(I-\tilde{M})^{-1}$$ provided that the limit exists.
\end{corollary}
{\it Proof.} Observe that by definition, the spectral radius of $\tilde{M}$ is strictly less than 1. Then from Corollary 5.6.16 of~\cite{HJ90},
$I-\tilde{M}$ is invertible, and $(I-\tilde{M})^{-1}=\sum_{m=0}^{\infty} \tilde{M}^m$. The result thus follows from
Lemma~\ref{lem:limit}.
\hfill $\Box$

We also need a notion of matrix representation for  super-operators, which was investigated in~\cite{YYFD12}. To do this, we fix an orthonormal basis $\{|k\> : k\in K\}$ of $\h$.
\begin{definition}
Let $\a= \{ E_i : i\in I\}\in \SO(\h)$ be a super-operator. The matrix representation of $\a$ is defined as
$$M_\a =\sum_{i\in I} E_i\otimes E_i^*.$$
Here the complex conjugate is taken according to the orthonormal basis $\{|k\> : k\in K\}$. 
\end{definition}
It is easy to check that $M_\a$ is independent of the choice of orthonormal basis and the Kraus operators $E_i$. 

Furthermore, we can define the matrix representation for a matrix of super-operators.
Let $\t= (\a_{i,j})$ be an $m\times n$-matrix of  super-operators. Then the matrix representation of $\t$, denoted $M_\t$, is defined as the block matrix 
$$M_\t = 
\left(
\begin{array}{cccc}
  M_{\a_{1,1}}   & \dots & M_{\a_{1,n}}   \\
 \vdots & \ddots  & \vdots  \\
  M_{\a_{m,1}}   & \dots & M_{\a_{m,n}}   \\
\end{array}
\right)
$$
where for each $i$ and $j$, $M_{\a_{i,j}}$ is the matrix representation of $\a_{i,j}$.
In particular, let $\widetilde{\e}=(\a_1,\dots,\a_n)$ be a (row) vector of  super-operators. Then $M_{\widetilde{\e}}=(M_{\a_1},\dots,M_{\a_n}).$
We always denote by $M_X$ the matrix representation of $X$, whenever $X$ is a super-operator, a vector of super-operators, 
or a matrix of super-operators.

Let $|\Psi\>=\sum_{k\in K} |kk\>$ be a (un-normalised) maximally entangled state in $\h\otimes \h$. The next lemma is from~\cite{YYFD12}. 

\begin{lemma}\label{lem:corres}
Let $M_\a$ be the matrix representation of $\e\in \SO(\h)$. Then for any $E\in \lh$, we have
$$(\e(E) \otimes I_\h)|\Psi\> = M_\a(E \otimes I_\h)|\Psi\>.$$
\end{lemma}
Let $\e=\{ E_i : i\in I\}$ be a super-operator on $\h$. Then for any $k,l\in K$,
\begin{eqnarray}
\<k|\left(\sum_{i\in I} E_i^\dag E_i\right) |l\>
 &=& \tr \sum_{i\in I} E_i |l\>\<k| E_i^\dag\nonumber\\
& =& \tr [\a(|l\>\<k|)] \nonumber\\
&=& \<\Psi|\a(|l\>\<k|)\otimes I_\h |\Psi\>\label{eq:tmp161111}\\
& = &\<\Psi|M_\a(|l\>\<k|\otimes I_\h) |\Psi\>\nonumber\\
&=&\<\Psi|M_\a|lk\>.\nonumber
\end{eqnarray}
Note that for an operator $E\in \lh$, $\<k | E |l\>$ is exactly the $(k,l)$-th element of the matrix representation of $E$ under the basis $\{|k\> : k\in K\}$.
Lemma~\ref{lem:corres} indeed provides us an efficient way to determine whether or not $\e\lesssim\f$ when the matrix representations of $\e$ and $\f$ are given.

\subsubsection{Proof of Theorem~\ref{thm:fixpoint}.} We now turn to the proof of Theorem~\ref{thm:fixpoint}.
For the first part, we check that $f(X)$ indeed maps $\leqI^n$ into $\leqI^n$. Let $\widetilde{\f}\in \leqI^n$. Then 
for each $j$, 
\begin{eqnarray*}
f(\widetilde{\f})_j = \sum_{i=1}^n \widetilde{\f}_i \t_{i, j} + \widetilde{\g}_j\ \lesssim\  \sum_{i=1}^n \t_{i, j} + \widetilde{\g}_j\ \lesssim\  \id_\h
\end{eqnarray*}
where the first inequality is from Lemma~\ref{lem:rightapp} and the fact that $\widetilde{\f}_i\lesssim \id_\h$.
Note that the function $f$ defined in~Eq.(\ref{eq:les}) is Scott continuous with respect to the partial order $\le$. Then by Lemma~\ref{lem:cpov} and Kleene fixed point theorem, $f(X)$ has a (unique) least fixed point which can be written as $$\widetilde{\e} =  \sum_{m=0}^\infty \widetilde{\g}\t^m.$$

\begin{table}[t]
\begin{lrbox}{\tablebox}
\begin{tabular}{l}
\hline
\begin{tabular}{l}
Input: 
(1) An $n\times n$ matrix  $\t=(\t_{i,j})$ and a $1\times n$ vector $\widetilde{\g}$ of super-operators such that for each $j$,\\
\qquad\qquad\ \  $\sum_i \t_{i,j} + \widetilde{\g}_j \lesssim \id_\h$.\\
\ \qquad \ \ (2) A super-operator $\e\in \leqI$ and $i$ such that $1\leq i\leq n$.\\
Output: `yes' if $\e\sim \widetilde{\e}_i$ and `no' otherwise, where $\widetilde{\e}$ is the least fixed point of Eq.(\ref{eq:les}).\\
\hline
(1) Construct $M_{\widetilde{\g}}$ and $M_{\t}$ from $\widetilde{\g}$ and $\t$, respectively;\\
(2) Calculate the Jordan decomposition $M_{\t}=S J S^{-1}$; \\
(3) Compute $M_{\widetilde{\e}}=M_{\widetilde{\g}}S (I-\tilde{J})^{-1} S^{-1}$;\\
(4) Use the method described in Eq.(\ref{eq:tmp161111}) to determine if $\widetilde{\e}\sim  \widetilde{\e}_i$, and output the result. \\
\hline
\end{tabular}
\end{tabular}
\end{lrbox}
\resizebox{\textwidth}{!}{\usebox{\tablebox}}
\caption{Algorithm for Theorem~\ref{thm:fixpoint}.}\label{tb:thm1al}
\end{table}%

For the second part of Theorem~\ref{thm:fixpoint}, we first prove that the matrix representation of $\widetilde{\e}$ is 
\begin{equation}\label{eq:tmp12301}
M_{\widetilde{\e}}=M_{\widetilde{\g}}(I-\tilde{M}_\t)^{-1}.
\end{equation}
For any $1\leq i\leq n$ and $E_i\in \lh$, we calculate from Lemma~\ref{lem:corres} that
\begin{eqnarray*}
M_{\widetilde{\e}_i} (E_i\otimes I_\h) |\Psi\>
&=& \widetilde{\e}_i(E_i)\otimes I_\h |\Psi\>\\
&=&\sum_{m=0}^\infty \sum_{k_1,\dots, k_m=1}^n \widetilde{\g}_{k_1}\t_{k_1,k_2}\cdots \t_{k_{m},i}(E_i)\otimes I_\h|\Psi\>\\
&=&\sum_{m=0}^\infty \sum_{k_1,\dots, k_m=1}^n (M_{\widetilde{\g}_{k_1}}M_{\t_{k_1,k_2}}\cdots M_{\t_{k_{m},i}})(E_i\otimes I_\h)|\Psi\>\\
&=&\sum_{m=0}^\infty (M_{\widetilde{\g}} M_\t^m)_i(E_i\otimes I_\h)|\Psi\>\\
\end{eqnarray*}
where $(M_{\widetilde{\g}} M_\t^m)_i$ is the $i$-th block of $M_{\widetilde{\g}} M_\t^m$ at the corresponding position of $M_{\widetilde{\e}_i}$ in $M_{\widetilde{\e}}$. Note that when $E_i$ ranges over $\lh$, the state $(E_i\otimes I_\h) |\Psi\>$ ranges over all pure state in $\h\otimes \h$. The above equations indeed imply that $M_{\widetilde{\e}_i}=\sum_{m=0}^\infty (M_{\widetilde{\g}} M_\t^m)_i$, and thus $M_{\widetilde{\e}}=\sum_{m=0}^\infty M_{\widetilde{\g}} M_\t^m$. Then Eq.(\ref{eq:tmp12301}) follows from Corollary~\ref{cor:limit}. 

An algorithm which determines for any $\e\in \leqI$ and $1\leq i\leq n$ if $\e\sim \widetilde{\e}_i$ is sketched 
in Table~\ref{tb:thm1al}.
It is easy to see that the time complexity is $c n^2$, where the constant $c$ is of the order $d^4$, and $d$ is the dimension of the Hilbert space $\h$. \qed
\subsection{Proof of Theorem~\ref{thm:mainthm}}\label{sec:pmain}

\subsubsection{Basic results for vector measures.}

We review some necessary definitions and results for vector measures. For more details, please refer to~\cite{DU77}.

Let $\Omega$ be a non-empty set. A semi-algebra $\s$ on $\Omega$ is a subset  of the power set $2^\Omega$ with the following properties:
\begin{enumerate}
\item $\emptyset\in \s$;
\item $A, B\in \s$ implies $A\cap B\in \s$;
\item $A, B\in \s$ implies that $A\backslash B=\biguplus_{i=1}^n A_i$ for some disjoint $A_1, \dots, A_n \in \s$.
\end{enumerate}

An algebra is a semi-algebra which is further closed under union and subtraction; a $\sigma$-algebra is an algebra which is also closed under complement and countable union. Given a semi-algebra $\s$, we denote by $\r(\s)$ (resp. $\sigma(\s)$) the algebra (resp. $\sigma$-algebra)  generated by
$\s$; that is, the intersection of all algebras  (resp. $\sigma$-algebras) which contain $\s$ as a subset. Obviously, $\sigma(\s) = \sigma(\r(\s))$.

Recall also that a Banach space is a complete normed vector space. 
\begin{definition}
Let $T\subseteq 2^\Omega$, and 
 $\fdmu$ a function from $T$ to a Banach space $\mathcal{B}$. We call $\fdmu$ a countably additive vector measure, or vector measure for simplicity, if  for any sequence $(A_i)_{i\geq 1}$ of pairwise disjoint members of $T$ such that $\biguplus_{i=1}^\infty A_i\in T$, it holds that
 $$\fdmu(\biguplus_{i=1}^\infty A_i) = \sum_{i=1}^\infty \fdmu(A_i).$$
\end{definition}

\begin{definition}
Let $\r$ be an algebra on $\Omega$ and $\fdmu : \r \ra \mathcal{B}$ a vector measure. Let $\mu$ be a finite nonnegative real-valued measure
on $\r$. Then $\fdmu$ is said to be $\mu$-continuous if $$\lim_{\mu(A)\ra 0} \fdmu(A) =0.$$
\end{definition}

The next theorem from~\cite{DU77} is the key to prove Theorem~\ref{thm:mainthm}.
\begin{theorem}\label{thm:cara}{\rm [Carath\'eodory-Hahn-Kluvanek Extension Theorem]}
Let $\r$ be an algebra on $\Omega$ and $\fdmu : \r\ra \mathcal{B}$ a bounded vector measure. If there exists a finite and nonnegative real-valued measure $\mu$ on $\r$ such that $\fdmu$ is $\mu$-continuous, then $\fdmu$ can be uniquely extended to a vector measure $\fdmu' : \sigma(\r) \ra \mathcal{B}$ on the $\sigma$-algebra generated by $\r$ such that
$$\fdmu'(A) = \fdmu(A) \mbox{ for any } A\in \r.$$
\end{theorem}

\subsubsection{Banach space of Hermitian-preserving mappings.}

Suppose $\h$ is a Hilbert space. Let $\HP(\h)$ be the set of Hermitian-preserving linear operators on $\lh$; that is
$$\HP(\h) = \{\hpe \in \mathcal{L}(\lh)\ |\ \hpe(E)\mbox{ is Hermitian provided that $E$ is Hermitian}\}.$$
It is easy to show that $\hpe$ is Hermitian-preserving if and only if for any $E\in \lh$, $\e(E^\dag) = \e(E)^\dag$.
Obviously, $(\HP(\h), +,  \circ)$ forms a ring, $\SO(\h) \subseteq \HP(\h)$, and the orders $\le$ and $\lesssim$ defined in Definition~\ref{def:orders} can be lifted to $\HP(\h)$.
We denote by $\HP_\eqsim(\h)$ the quotient set of $\HP(\h)$ by $\eqsim$, and $[\e]\in \HP_\eqsim(\h)$ the equivalent class 
of $\e\in \HP(\h)$. Note that although the quotient set $\HP_\eqsim(\h)$ is again an Abelian group with respect to the addition $+$, by defining $[\e] + [\f] = [\e+\f]$, the composition $\circ$
is not even well defined in $\HP_\eqsim(\h)$: Let 
$\e=\{X\}$ be the Pauli-$X$ super-operator, and $\c=\{|0\>\<0|\}$. Then $[\a]= [\id_\h]$, but $\c\a\not\eqsim \c\id_\h$ since $\tr(\c\a(|0\>\<0|))=0$ while  $\tr(\c\id_\h(|0\>\<0|))=1$.

 We further extend the pre-order $\lesssim$ to the quotient set $\HP_\eqsim(\h)$ by letting
$[\e]\lesssim [\f]$ if $\e\lesssim \f$. It is easy to check that this definition is well defined, and $\lesssim$ becomes a partial order in $\HP_\eqsim(\h)$.

\begin{lemma}Let $\|\cdot\|$ be an arbitrary operator norm. Then
the quotient set $(\HP_\eqsim(\h), \|\cdot\|_\eqsim)$ is again a Banach space, where the norm $\|\cdot\|_\eqsim$ is defined by
$$\|[\hpe]\|_\eqsim = \inf_{\f\in [\hpe]}\|\f\|.$$
\end{lemma} 
{\it Proof.}
Note that $\HP(\h)$ is a finite dimensional vector space over the field of real numbers. Thus $(\HP, \|\cdot\|)$ is a Banach space, and so is the quotient space 
$(\HP_\eqsim(\h), \|\cdot\|_\eqsim)$.
\hfill $\Box$

\subsubsection{Proof of Theorem~\ref{thm:mainthm}.}
We are now ready to prove Theorem~\ref{thm:mainthm}. To do this, we first regard the mapping $Q_s$ defined in Eq.(\ref{eq:qs}) as from $\s^\m(s)$ to $\HP(\h)$. Let $Q_s'$ be a mapping from
$\s^\m(s)$ to $\HP_\eqsim(\h)$ such that for any $A\in \s^\m(s)$, $$Q_s'(A)=[Q_s(A)].$$ 
Then we have the following lemmas.
\begin{lemma}
The mapping $Q'_s$ defined above is a bounded vector measure over $\s^\m(s)$.
\end{lemma}
{\it Proof.}
We only need to check that $Q'_s$ is countably additive. Let $\emptyset\neq A=\biguplus_{i\geq 1} A_i$ for a disjoint sequence $(A_i)_{i\geq 1}$ in $\s^\m(s)$ and $A\in \s^\m(s)$. We need to show
\begin{equation}\label{eq:tmp0124}
Q'_s(A) = \sum_{i\geq 1} Q'_s(A_i). 
\end{equation}
Without loss of generality, we assume that the sequence $(A_i)_{i\geq 1}$ has been arranged such that all empty sets, if there are any, are put at the tail of the sequence; that is, whenever $A_n\neq \emptyset$ then $A_i\neq \emptyset$ for any $i\leq n$.

We claim that there are only finitely many nonempty sets in the sequence; that is, there exists $n$ such that $A_i\neq \emptyset$ if and only if $i\leq n$. We prove this claim by contradiction. Suppose $A=Cyl(\widehat{\pi}_0)$, and for each $i\geq 1$, $A_i=Cyl(\widehat{\pi}_i)$ where all $\widehat{\pi}_i$s are in $\mathit{Path_{fin}^{\m}(s)}$ and $\widehat{\pi}_0$ is a prefix of each $\widehat{\pi}_i$ for $i\geq 1$. By the fact that $A_i$s are disjoint, $\widehat{\pi}_i$ cannot be a prefix of $\widehat{\pi}_j$ for distinct $i, j\geq 1$. Let $\Pi=\{\widehat{\pi}_i : i\geq 1\}$. For any $\widehat{\pi}\in \mathit{Path_{fin}^{\m}(s)}$, let $Ind_{\widehat{\pi}} = \{ i\geq 1 : \widehat{\pi} \mbox{ is a prefix of } \widehat{\pi}_i\}$, and
$$K= \{ \widehat{\pi}\in \mathit{Path_{fin}^{\m}(s)} : Ind_{\widehat{\pi}} \mbox{ is infinite}\}.$$ 
Obviously, $K\cap \Pi =\emptyset$. Note that $\widehat{\pi}_0\in K$, and for any $\widehat{\pi}\in K$, since 
$$Ind_{\widehat{\pi}}=\biguplus_{t\in S}Ind_{\widehat{\pi}^\frown t},$$
there exists $t_{\widehat{\pi}}\in S$ such that $\widehat{\pi}^\frown t_{\widehat{\pi}}\in K$ again. Thus we can extend $\widehat{\pi}_0$ to an (infinite) path $\pi\in \mathit{Path^{\m}(s)}$ such that any finite-length prefix $\widehat{\pi}$ of $\pi$ with $|\widehat{\pi}|\geq |\widehat{\pi}_0|$ is not included in $\Pi$. Thus for any $i$, $\pi\not \in A_i$, contradicting the fact that $\pi\in A$.

With the claim in hand, we let $N=\max\{|\widehat{\pi}_i|: 1\leq i\leq n\}$
and  $\Pi_{N} = \{\widehat{\pi}\in \Pi : |\widehat{\pi}|= N\}$. Obviously, we can partition $\Pi_N$ into several disjoint subsets such that each one contains exactly $|S|$ elements with the same $N-1$-length prefix; that is, there exists a set $\{\widehat{\pi}'_1, \dots, \widehat{\pi}'_{I_N}\}$ such that for each $1\leq i\leq I_N$, 
$|\widehat{\pi}'_i|=N-1$, and 
$$\Pi_N = \biguplus_{i=1}^{I_{N}} \Pi_N^i\mbox{ where } \Pi_N^i = \{\widehat{\pi}_i'{}^\frown  t: t\in S\}.$$
We delete $\Pi_{N}$ from $\Pi$ and add $\widehat{\pi}'_i$, $1\leq i\leq I_N$, into it. Denote by $\Pi_{\leq N-1}$ the resultant set. Then 
each element in $\Pi_{\leq N-1}$ has length less than $N$, and it is easy to check that
$$\sum_{\widehat{\pi}\in \Pi_N} \q(\widehat{\pi})\eqsim \sum_{i=1}^{I_N} \q(\widehat{\pi}_i'), \mbox{ thus } \sum_{\widehat{\pi}\in \Pi} \q(\widehat{\pi})\eqsim \sum_{\widehat{\pi}\in \Pi_{\leq N-1}} \q(\widehat{\pi}).$$
Proceeding in this way, we can construct a sequence of sets $\Pi_i, |\widehat{\pi}_0|+1\leq i\leq N,$ such that for any $i$,
$\sum_{\widehat{\pi}\in \Pi_{\leq i}} \q(\widehat{\pi})\eqsim \sum_{\widehat{\pi}\in \Pi_{\leq i-1}} \q(\widehat{\pi})$ where $\Pi_{\leq N} = \Pi$.
Note that $\Pi_{\leq |\widehat{\pi}_0|}=\{\widehat{\pi}_0\}$. We finally have 
$$\sum_{i\geq 1} Q'_s(A_i)=\sum_{\widehat{\pi}\in \Pi} [\q(\widehat{\pi})] = [\q(\widehat{\pi}_0)]=Q'_s(A).$$
That completes the proof of the lemma.\hfill $\Box$
\begin{lemma} The mapping $Q'_s$ defined above can be extended uniquely to a vector measure, denoted by $Q'_s$ again, from $\sigma(\s^\m(s))$ to $\HP_{\eqsim}(\h)$. Furthermore, for any $A\in \sigma(\s^\m(s))$, 
\begin{equation}\label{eq:tmp1230}
[0_\h]\lesssim Q'_s(A) \lesssim [\id_\h].
\end{equation}
\end{lemma}
{\it Proof.} Let $\r=\r(\s^\m(s))$ be the algebra generated by $\s^\m(s)$. Obviously, we have
$$\r = \{ A : A=\biguplus_{i=1}^n A_i \mbox{ for some } n\geq 0, A_i\in \s^\m(s)\}.$$
We extend the mapping $Q'_s$ to $\r$ by defining $Q'_s(\biguplus_{i=1}^n A_i) = \sum_{i=1}^nQ'_s(A_i)$, which turns out to be a bounded vector measure from $\r$ to $\HP_\eqsim(\h)$.
Let $\mu_s$ be a mapping defined as follows:
\begin{itemize}
\item $\mu_s(\emptyset)=0$, and for any $A=Cyl(\widehat{\pi})\in \s^\m(s)$, $\mu_s(A) = \tr(\q(\widehat{\pi})(\rho_m))$ where $\rho_m = I_{dim(\h)}/dim(\h)$ is the maximally mixed state in $\dh$;
\item for any disjoint sets $A_1, \dots, A_n$ in $\s^\m(s)$,  
$\mu_s(\biguplus_{i=1}^n A_i)=\sum_{i=1}^n \mu_s(A_i)$.
\end{itemize}
Then $\mu_s$ is indeed a finite and nonnegative real-valued measure on $\r$, since $\mu_s(\mathit{Path^{\m}(s)}) =\mu_s(Cyl(s)) =  \tr(\id_\h(\rho_m)) = 1$.
Note that for any super-operator $\e=\{E_i : i\in I\}$, $\tr(\e(\rho_m))=\sum_{i\in I} \| E_i\|_2^2/dim(\h)$ where $\|\cdot\|_2$ is the Euclidean norm.
It follows that if  $\lim_{i\ra \infty} \tr(\e_i(\rho_m))=0$, where $(\e_i)_{i\geq 1}$ is a sequence of super-operators, then $\lim_{i\ra \infty} [\e_i]=[0_\h]$.
So we have $\lim_{\mu_s(A)\ra 0} Q'_s(A) = [0_\h]$, which means that $Q'_s$ is $\mu_s$-continuous.

Now using Theorem~\ref{thm:cara}, we can extend $Q'_s$ uniquely to a vector measure $Q'_s : \sigma(\s^\m(s)) \ra \HP_\eqsim(\h)$. In the following, we show that the
extension satisfies Eq.(\ref{eq:tmp1230}). By the additivity of $Q'_s$, it suffices to show that  for any $A\in \sigma(\s^\m(s))$, 
 $[0_\h]\lesssim Q'_s(A)$; that is, for any $\rho\in \dh$, $\tr(Q'_s(A)(\rho))\geq 0$. Let $\mu_{\rho} : \sigma(\s^\m(s)) \ra \sreal$ be defined as 
$$\forall A\in \sigma(\s^\m(s)): \mu_{\rho}(A) =\tr(Q'_s(A)(\rho)).$$
Then obviously, $\mu_{\rho}$ is an (ordinary) real-valued measure over  $\sigma(\s^\m(s))$ and its restriction 
on $\s^\m(s)$, denoted $\mu_{\rho}|_{\s^\m(s)}$, is a probabilistic measure.
Now from Carath\'eodory Theorem for probabilistic measures~\cite{Bi08}, $\mu_{\rho}|_{\s^\m(s)}$ can be uniquely extended to a probabilistic measure $\mu_{\rho}'$ over $\sigma(\s^\m(s))$. Then we have $\tr(Q'_s(A)(\rho))=\mu_{\rho}(A) =\mu_{\rho}'(A)\geq 0$ by the uniqueness of such extension.
\hfill $\Box$

With the two lemmas above, we can easily prove Theorem~\ref{thm:mainthm}. For any $A\in \sigma(\s^\m(s))$, let $Q_s(A)$ be any super-operator in the equivalent class $Q_s'(A)$. It is obvious that such an extension is indeed a SVM, and it is unique up to the equivalence relation $\eqsim$. \qed
\end{document}